\documentclass[letterpaper,12pt]{article}
\usepackage[pdftex]{hyperref}
\usepackage{graphicx}
\usepackage{fullpage}

\begin{document}
%%%%%%%%%%%%%%%%%%%%%%%%%%%%%%%%%%%%%%%%%%%

\def\a{\alpha}
\def\b{\beta}
\def\c{\varepsilon}
\def\d{\delta}
\def\e{\epsilon}
\def\f{\phi}
\def\g{\gamma}
\def\h{\theta}
\def\k{\kappa}
\def\l{\lambda}
\def\m{\mu}
\def\n{\nu}
\def\p{\psi}
\def\q{\partial}
\def\r{\rho}
\def\s{\sigma}
\def\t{\tau}
\def\u{\upsilon}
\def\v{\varphi}
\def\w{\omega}
\def\x{\xi}
\def\y{\eta}
\def\z{\zeta}
\def\D{\Delta}
\def\G{\Gamma}
\def\H{\Theta}
\def\L{\Lambda}
\def\F{\Phi}
\def\P{\Psi}
\def\S{\Sigma}

\def\o{\over}
\def\beq{\begin{eqnarray}}
\def\eeq{\end{eqnarray}}
\newcommand{\gsim}{ \mathop{}_{\textstyle \sim}^{\textstyle >} }
\newcommand{\lsim}{ \mathop{}_{\textstyle \sim}^{\textstyle <} }
\newcommand{\vev}[1]{ \left\langle {#1} \right\rangle }
\newcommand{\bra}[1]{ \langle {#1} | }
\newcommand{\ket}[1]{ | {#1} \rangle }
\newcommand{\EV}{ {\rm eV} }
\newcommand{\KEV}{ {\rm keV} }
\newcommand{\MEV}{ {\rm MeV} }
\newcommand{\GEV}{ {\rm GeV} }
\newcommand{\TEV}{ {\rm TeV} }
\def\diag{\mathop{\rm diag}\nolimits}
\def\Spin{\mathop{\rm Spin}}
\def\SO{\mathop{\rm SO}}
\def\O{\mathop{\rm O}}
\def\SU{\mathop{\rm SU}}
\def\U{\mathop{\rm U}}
\def\Sp{\mathop{\rm Sp}}
\def\SL{\mathop{\rm SL}}
\def\tr{\mathop{\rm tr}}

\def\IJMP{Int.~J.~Mod.~Phys. }
\def\MPL{Mod.~Phys.~Lett. }
\def\NP{Nucl.~Phys. }
\def\PL{Phys.~Lett. }
\def\PR{Phys.~Rev. }
\def\PRL{Phys.~Rev.~Lett. }
\def\PTP{Prog.~Theor.~Phys. }
\def\ZP{Z.~Phys. }

%%%%%%%%%%%%%%%%%%%%%%%%%%%%%%%%%%%%%%%%%%%%%%%%%%%%%%%%%%%%%%%%%%%%

\baselineskip 0.7cm
%\widowpenalty=1500
%\clubpenalty=1500

\begin{titlepage}

\begin{flushright}
IPMU10-0071
\end{flushright}

\vskip 1.35cm
\begin{center}
{\large \bf
  Testing the Nambu-Goldstone Hypothesis for Quarks and Leptons at the LHC
}
\vskip 1.2cm
Sourav K. Mandal$^{1,2}$, Mihoko Nojiri$^{2,3,4}$, Matthew Sudano$^{2,5}$ and  Tsutomu T. Yanagida$^{2,6}$
\vskip 0.4cm

{\it  
$^1$Department of Physics, University of California\\
Berkeley, CA 94720, USA\\
$^2$Institute for the Physics and Mathematics of the Universe, University of Tokyo\\
Kashiwa, Chiba 277-8583, Japan\\
$^3$Theory Group, KEK, %1-1 Oho, 
Tsukuba, Ibaraki 305-0801, Japan\\
$^4$The Graduate University for Advanced Studies (SOKENDAI)\\
Tsukuba, Ibaraki 305-0801, Japan\\
$^5$School of Natural Sciences, Institute for Advanced Study\\
Princeton, NJ 08540, USA\\
$^6$Department of Physics, Graduate School of Science, University of Tokyo\\
Tokyo 113-0033, Japan
}

\vskip 1.5cm

\abstract{
The hierarchy of the Yukawa couplings is an outstanding problem of the standard model.  We present a class of models in which the first and second generation fermions are SUSY partners of pseudo-Nambu-Goldstone bosons that parameterize a non-compact K\"ahler manifold, explaining the small values of these fermion masses relative to those of the third generation.  We also provide an example of such a model.  We find that various regions of the parameter space in this scenario can give the correct dark matter abundance, and that nearly all of these regions evade other phenomenological constraints.  We show that for $m_{\tilde g}\sim 700$~GeV, model points from these regions can be easily distinguished from other mSUGRA points at the LHC with only $7\;{\rm fb}^{-1}$ of integrated luminosity at 14~TeV.  The most striking signatures are a dearth of $b$- and $\tau$-jets, a great number of multi-lepton events, and either an ``inverted'' slepton mass hierarchy, narrowed slepton mass hierarchy, or characteristic small-$\mu$ spectrum.
}
\end{center}
\end{titlepage}

\setcounter{page}{2}
%\tableofcontents

\section{Introduction}

One of the fundamental problems in particle physics is to explain the hierarchy in the Yukawa couplings of the quarks and leptons. This is a long-standing problem, and in fact many models have been proposed to account for the smallness of the Yukawa couplings of the first and second generations relative to those of the third. In supersymmetric (SUSY) theories, there arises an intriguing possibility that the quarks and leptons are nothing but SUSY partners of Nambu-Goldstone (NG) bosons~\cite{BLPY,BPY}, where the Yukawa couplings are forbidden by the celebrated low-energy theorem~\cite{Adler}. Thus, the small Yukawa couplings for the first and second generations are regarded as small breakings of postulated symmetries.

If that is indeed the case, then squarks and sleptons in the first and second generations are approximately massless at some cut-off scale $\Lambda$ of the theory and they acquire masses from radiative corrections. 
If the corrections are dominated by the standard model gauge interactions, then they each have nearly flavor-independent masses, solving the SUSY flavor-changing neutral current problem. This hypothesis predicts a remarkable spectrum for SUSY particles at the electroweak scale. The purpose of this paper is to examine the low-energy implications of the Nambu-Goldstone hypothesis.

It is very important to note here that if the K\"ahler manifold parameterized by the NG bosons is a compact manifold such as ${\bf CP^n}$, the symmetry is explicitly broken by a constant term ${\cal O}(m_{3/2})$ in the superpotential~\cite{BW}\footnote{It is very interesting that two light generations in an $SO(10)$ GUT are naturally accommodated in $E_7/SO(10)\times U(1)^2$. It has been shown that if one eliminates the $U(1)^2$ in the unbroken subgroup one can couple the non-linear sigma model to supergravity without any explicit breaking of the $E_7$~\cite{Kugo-Yanagida}. We will discuss the $E_7$ model in a separate paper.}. As a consequence of the explicit breaking, the NG bosons have masses of ${\cal O}(m_{3/2})$, and it will be not easy to test the NG hypothesis at the LHC. Thus, we consider in this paper some non-compact complex manifold (non-linear sigma model) accommodating squarks and sleptons in the first and second generations as massless NG bosons. (The results in this paper do not depend on the explicit model for the non-compact complex manifold. We show it for the existence proof of a model.) We treat quark and lepton chiral multiplets in the third generation and Higgs chiral multiplets as fundamental fields and hence their scalar bosons have soft SUSY-breaking masses of order of the gravitino mass, $m_{3/2},$ at the cut-off scale $\Lambda$ \footnote{The boundary condition of SUSY-breaking masses in the present paper may be also realized in a brane world. That is, one assumes that quark and lepton multiplets in the first and second generation are confined on one brane separated from the other brane on which the third-generation quark and lepton, Higgs and hidden-sector multiplets reside. The gauge multiplets are in the bulk.  This model may generate a partially realized Higgs-exempt~\cite{jason} no-scale type supergravity model~\cite{EN} where squarks and sleptons in the first and second generation have very small SUSY-breaking masses, while those in the third generation have masses of order the gravitino mass $m_{3/2}$.}.

\section{An example of a non-compact manifold for the first and second generations}

In defining a supersymmetric non-linear sigma model, it is not sufficient to specify a symmetry breaking $G\rightarrow H$.  While this determines the number of Nambu-Goldstone (NG) bosons,
the number of NG chiral multiplets is not uniquely determined \cite{KOY,BKMU}.  Consider the simple example of  the $SU(2)/U(1)$ non-linear sigma model. This manifold is parameterized by two massless NG bosons, $x_1$ and $x_2$. We may introduce two chiral multiplets, $\phi_1$ and $\phi_2$, each of which contains one NG boson and one additional real, massless scalar. This realization is referred to as ``doubled.''  In this scenario, there are two charge eigenstates, 
\begin{equation}
\phi_+ = \phi_1+\phi_2,\qquad\phi_-=\phi_1-\phi_2,
\end{equation}
which form a real representation of the unbroken subgroup.  This is a general and unacceptable feature of the doubled realization because the standard model is chiral.  

In the fully non-doubled or ``pure'' realization, the two true NG bosons lie in a single NG chiral multiplet, $\phi^+$.  In this case, the $\phi^-$ is absent and the NG chiral multiplet $\phi^+$ is a complex representation of the unbroken $U(1)$ subgroup. The resultant manifold is nothing but $CP^1$. Unfortunately, it became known that the pure realization cannot be coupled to supergravity~\cite{BW}. However, it has been recently pointed out that a hybrid realization with some doubled and some non-doubled NG multiplets does not suffer from this inconsistency~\cite{Kugo-Yanagida}. We will therefore search for a hybrid non-linear sigma model that contains two generations of quarks and leptons as NG chiral multiplets. 

Let us first consider a non-compact complex manifold that accommodates one generation of left-handed quark and lepton multiplets. In particular, take a SUSY $U(6)/[U(4)\times SU(2)]$ non-linear sigma model which consists of $(4\times 2 +1)$ NG chiral multiplets, $\phi_i^a$ and $\varphi$, where $a=1,\dots,4$ and $i=1,2$.  The former superfield contains the left-handed quarks and leptons, 
\begin{equation}
\phi^a_i=\left(
\begin{array}{cc}
u_L^\xi & \nu_L \\
d_L^\xi & e^-_L
\end{array}
\right)\sim({\bf 4},{\bf 2})
\end{equation}
where $\xi=1,2,3$ is the color index.
The latter is a non-standard-model field known as the ``novino''~\cite{BPY}. All the bosons in $\phi^a_i$ and one real boson in $\varphi$ are NG-boson coordinates of  the $U(6)/[U(4)\times SU(2)]$ manifold. The other extra real boson in the novino superfield, $\varphi$, is necessary to construct the SUSY $U(6)/[U(4)\times SU(2)]$ non-linear sigma model~\cite{KOY}. Then, we have a complex manifold which consists of all 18 bosons.

We arrange these into a matrix,
\begin{equation}
\Psi= \left(
\begin{array}{c}
e^{\kappa \varphi/v}{\bf 1}_2\\
\phi^a_i/v
\end{array}
\right),
\end{equation}
which has 12 components $\Psi^\alpha_i$, $\alpha=1,\dots,6$. The general form of the K\"ahler potential of the SUSY $U(6)/[U(4)\times SU(2)]$ non-linear sigma model is then given by~\cite{KUY}
\begin{equation}
\label{lag}
K=v^2F(\rm{det}[\Psi^{\dagger }_\alpha \Psi^\alpha ]),
\end{equation}
where the dimension-one constant $v$ is determined, together with the constant $\kappa$, by normalization conditions for the NG chiral multiplets as 

\begin{equation}\label{d2f1}
v^2\frac{\partial^2 F}{\partial \phi^{\dagger i}_a\partial \phi^b_j} \bigg|_{\phi^a_i =\varphi=0} =\delta^a_b \delta^j_i
\end{equation}
\begin{equation}\label{d2f2}
(\kappa v)^2\frac{\partial^2 F}{\partial \varphi^\dagger\partial \varphi} \bigg|_{\phi^a_i =\varphi=0} = 1.
\end{equation}
The parameter $v$ corresponds to the energy scale of the $U(6)$ breaking. Here, notice that the function $F(x)$ is an arbitrary function satisfying (\ref{d2f1}) and (\ref{d2f2}). This freedom originates from the presence of one extra (non-NG) boson in the novino superfield, $\varphi$~\cite{BPY, BKMU}.

The above K\"ahler potential is invariant under the $U(6)_{\rm global}\times SU(2)_{\rm local}$ symmetry
\begin{equation}
\Psi \rightarrow g\Psi h^{-1}(x,\theta),
\end{equation}
where $g$ is a parameter of the global $U(6)$ transformation and $h(x,\theta )$ is a chiral superfield parameter of the hidden local $SU(2)$ transformation. The form in (\ref{lag}) is maintained by using the local $SU(2)$ transformation. Thus, the global $U(6)$ symmetry is non-linearly realized by the NG chiral multiplets, $\phi^a_i$ and $\varphi$.

Let us now couple the above model to supergravity. The interaction is given by
\begin{equation}
\Delta{\cal L} = 3\int d^2\theta d^2{\bar \theta}{\cal E} {\rm exp}\Big(\frac{1}{3} K(\phi^a_i,\varphi,\phi^{\dagger i}_a,\varphi^\dagger)\Big).
\end{equation}
Here, ${\cal E}$ is the supervierbein determinant. It should be stressed here that the supergravity Lagrangian is completely invariant under the global $U(6)$ symmetry.

It is straightforward to introduce the hidden sector responsible for the SUSY breaking.  For simplicity, we introduce a single singlet field, $Z$, for this purpose. Then, the total K\"ahler potential is 
\begin{equation}
K= K(\phi^a_i,\varphi,\phi^{\dagger i}_a,\varphi^\dagger) + Z^\dagger Z + ....
\end{equation}
The introduction of this SUSY-breaking sector preserves the global $U(6)$ symmetry and, therefore, leaves our NG bosons massless. However, there is a real scalar in $\varphi$ that is not a NG boson, and it acquires a soft mass of ${\cal O}(m_{3/2})$. This fact is shown by an explicit calculation~\cite{GY}.

So far, only the left-handed quarks and leptons have been introduced in our non-linear sigma model. It is straightforward to accommodate the right-handed quarks and leptons in another $U(6)/[U(4)\times SU(2)]$ manifold. The NG chiral multiplets are
\begin{equation}
{\widetilde \Psi}=\left(
\begin{array}{c}
e^{\kappa {\widetilde \varphi}/v}{\bf1}_2\\
{\widetilde \phi^i_a/v} 
\end{array}
\right),
\end{equation}
where the ${\widetilde \phi}^i_a$ are the chiral multiplets for the right-handed quarks and leptons and the ${\widetilde \varphi}$ is another novino. The total manifold is now $(U(6)/[U(4)\times SU(2)])^2$.  The SM gauge group is a subgroup of the unbroken $(U(4)\times SU(2))^2$. If one gauges the $SU(2)\times SU(2)$ and a diagonal $SU(4)$ subgroup of the $U(4)\times U(4)$  in the unbroken symmetry, one obtains the Pati-Salam gauge model with one generation.

It is now clear how to extend the model to accommodate the second generation of quarks and leptons. That is, we consider a manifold $(U(10)/[U(8)\times SU(2)])^2$, such that the left and right multiplets are
\begin{eqnarray}
\left(\phi^a_i\right)_L &=& \left(
\begin{array}{cccc}
u_L^\xi & {\nu_e}_L & c_L^\xi & {\nu_\mu}_L\\
d_L^\xi & e^-_L & s_L^\xi & \mu^-_L\\
\end{array}
\right)\sim({\bf 8},{\bf 2})
\\
\left(\phi^a_i\right)_R &=& \left(
\begin{array}{cccc}
u_R^\xi & {\nu_e}_R & c_R^\xi & {\nu_\mu}_R\\
d_R^\xi & e^-_R & s_R^\xi & \mu^-_R\\
\end{array}
\right)\sim({\bf 8},{\bf 2})
\; .
\end{eqnarray}
Gauging suitable subgroups properly we obtain the supersymmetric standard model (SSM) with three generations, where the first two generations are Nambu-Goldstone bosons in the coset space and the third generation is fundamental.  After SUSY breaking, according to the low-energy theorem we have massless squarks and sleptons in the first two generations.  On the other hand, the Higgs and the squarks and sleptons in the  
third generation may have soft-SUSY breaking masses of ${\cal O}(m_ 
{3/2})$, since they are fundamental. The top Yukawa is also  
put in by hand, and given the non-degeneracy of the families, it breaks  
no symmetry and so is naturally taken to be ${\cal O}(1)$~\cite{tHooft}.

The absence of Yukawa interactions along with the masslessness of  
squarks and sleptons in the first and second generations is  
guaranteed when the global $U(10)\times U(10)$ symmetry is exact.  
However, once we introduce the SSM gauge interactions the global  
symmetry is explicitly broken and the squarks and sleptons are no  
longer true NG bosons.  Therefore, they may now appear in Yukawa  
interactions, but the Yukawa couplings are still suppressed.  This is because the NG boson description remains approximately valid, since the global symmetry is only weakly gauged.
After SUSY breaking, the radiative corrections from the gauge and  
Yukawa interactions induce small masses for the squarks and  
sleptons.

The induced masses are logarithmically divergent and hence we need counter terms in the K\"ahler potential. In principle, we cannot determine the counter terms, but we expect those terms to vanish at some cut-off scale $\Lambda$ where the present non-linear sigma model is realized. In the present paper, we simply assume the GUT scale $\simeq 2\times 10^{16}$ GeV to be the cut-off scale. In other words, we choose a boundary condition such that squarks and sleptons in the first and second generations are massless at the GUT scale. We easily see that squarks and sleptons in the first two generations have almost flavor-independent masses, since the SM gauge interactions are flavor blind and the Yukawa couplings are negligible compared with the gauge interactions. Thus, we do not have the serious flavor-changing neutral current problem.

However, this model does likely suffer from a gravitino problem~\cite{W} since $m_{3/2}$ is of ${\cal O}(1\;\rm{TeV})$.  The problem of late-decaying, non-LSP  gravitinos can be solved through a dilution process with a sufficiently low reheating temperature~\cite{GR}, but such a discussion is beyond the scope of this paper.  Similarly, although the novino may have interesting phenomenology because its mass is of ${\cal O}(m_{3/2})$, making numerical predictions would require an explicit model which we do not provide here.  
 
The results that follow do not rely on any particular model.  We assume that non-MSSM fields, including those of the GUT and the novino, are efficiently decoupled. Our results then follow from a set of SUSY-breaking parameters specified at the cut-off scale which are generic to the NG hypothesis on non-compact manifolds.

\section{Low-energy spectrum for SUSY particles}
\begin{figure}
\begin{center}
\includegraphics[width=0.9\columnwidth]{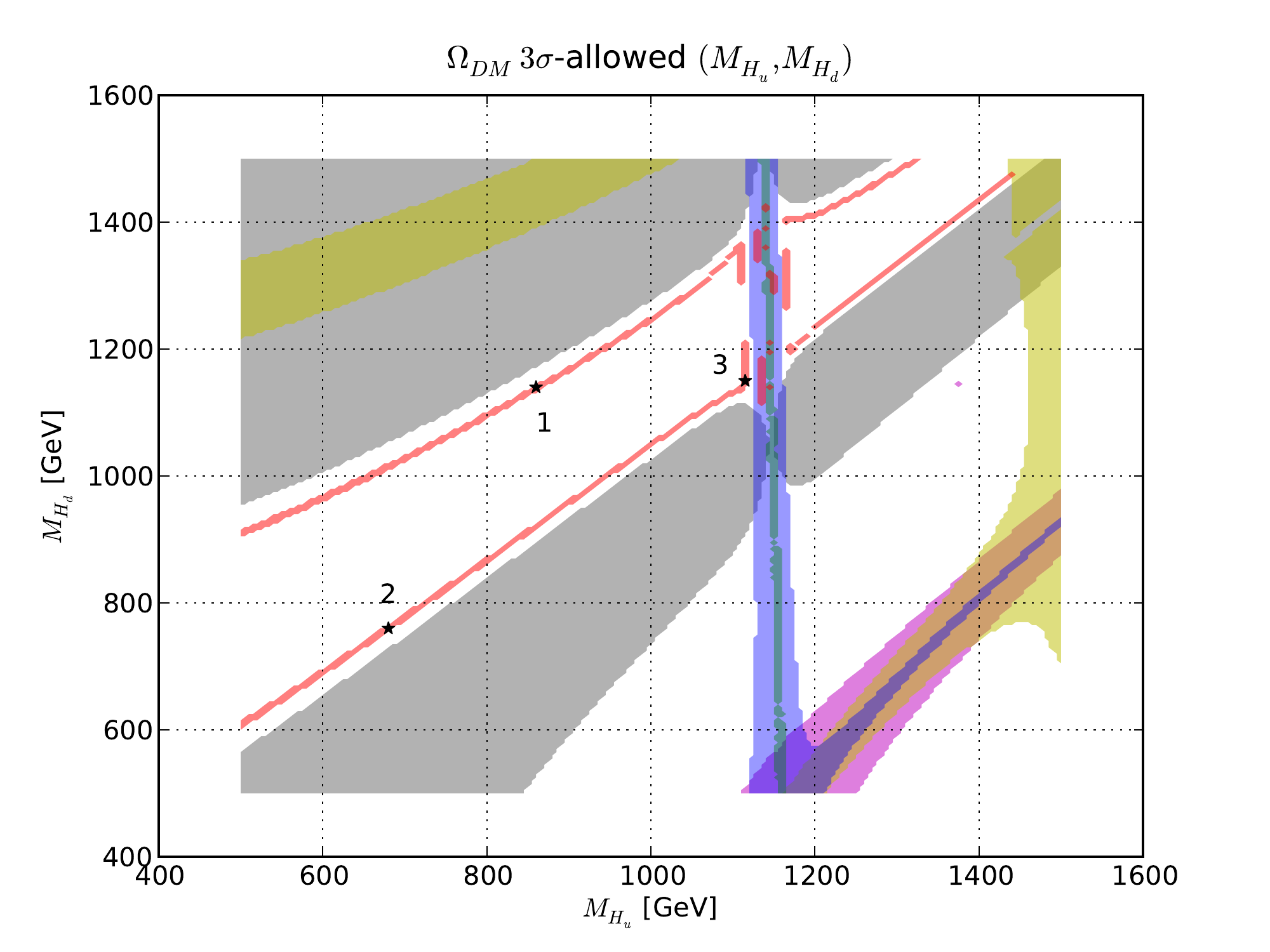}
\end{center}
\caption{Scan in $(M_{H_u}, M_{H_d})$ for the non-universal mSUGRA subspace $m_{1/2}=300\;\rm{GeV}$, $\tan\beta=10$, $\mu>0$ and $A_0=0$, with $m_0=0$ for the first and second generation scalars and $m_0=1\;\rm{TeV}$ for the third generation scalars.  The regions shown are: $3\sigma$-allowed dark matter relic density given by the seven-year WMAP data (red), charged LSP (gray), excluded by {\tt DarkSUSY} limits on $b\rightarrow s\gamma$ (magenta), excluded by {\tt DarkSUSY} Higgs mass bounds (yellow), excluded by 90\% limit on spin-independent WIMP-nucleon cross sections in direct detection experiments (blue), excluded by 90\% limit on spin-dependent WIMP-nucleon cross sections in direct detection experiments (green).  Benchmark points chosen for further study are denoted by the black stars (see text for details).}
\label{fig:dmscan}
\end{figure}

Consistent with the NG hypothesis on non-compact manifolds, we
consider the subspace of mSUGRA~\cite{CAN} models with the SUSY-breaking parameters
\begin{equation}
m_{1/2} = 300\;\rm{GeV},\quad \tan\beta=10,\quad \mu > 0,\quad A_0=0
\end{equation}
while setting $m_0=0$ for the first and second generation of scalars and $m_0=1\;\rm{TeV}$ for the third generation scalars.   Using the {\tt SOFTSUSY} SUSY spectrum calculator  (version 3.0.7)~\cite{SOFTSUSY} and the {\tt DarkSUSY} suite (version 5.0.5)~\cite{DarkSUSY,SLHA2,FeynHiggs,HiggsBounds} we scan the $(M_{H_u}, M_{H_d})$ parameter space where $M_{H_u}\sim M_{H_d}\sim{\cal O}(m_{3/2})$ to find the region which gives the correct dark matter relic density while also evading other phenomenological constraints.  We set the top mass to $m_t=175\;\rm{GeV}$ and let {\tt SOFTSUSY} solve for the GUT scale, which is always $\simeq 2\times 10^{16}$ GeV.
Results for the most relevant bounds are shown in Figure~\ref{fig:dmscan}.  The colored regions are as follows:
\begin{itemize}
\item {\bf red:} $3\sigma$-allowed dark matter relic density given by the seven-year WMAP data~\cite{WMAP7}.
\item {\bf gray:} Charged LSP.
\item {\bf magenta:} Excluded by {\tt DarkSUSY} limits on $b\rightarrow s\gamma$.
\item {\bf yellow:} Excluded by {\tt DarkSUSY} Higgs mass bounds.
\item {\bf blue:} Excluded by 90\% limit on spin-independent WIMP--nucleon cross sections in direct detection experiments~\cite{KSZ}.  
\item {\bf green:} Excluded by 90\% limit on spin-{\em dependent} WIMP--nucleon cross-sections in direct detection experiments~\cite{KSZ}.  For any given bin the more stringent limit among the proton or neutron cross-sections is chosen.
\end{itemize}
The allowed parameter space in Figure~\ref{fig:dmscan} shows some interesting properties.  First, significantly different combinations of $(M_{H_u},M_{H_d})$ can give the correct relic density.  In the upper and lower ``branch'' regions with a large difference between the two soft masses, coannihilation with $\tilde{e}$, $\tilde{\mu}$ or $\tilde{\nu}_{e,\mu}$ yields the required abundance.  The ``bridge'' at $M_{H_u}\approx M_{H_d}\approx 1100\;\rm{GeV}$ is a small-$\mu$ region where annihilation through $\tilde{h}$ contributes to the correct abundance for a relatively light neutralino~\cite{smallmu}.  Due to the large higgsino component in $\tilde\chi_1^0$, this bridge branch is near the region excluded by direct detection experiments (see, for example, Ref.~\cite{HNY}). For the upper, lower and bridge branches we have chosen benchmark points for further study, marked by black stars.  (These benchmark points give dark matter relic densities within $1\sigma$ of the WMAP7 value.)

\begin{figure}
\begin{center}
\includegraphics[width=0.9\columnwidth]{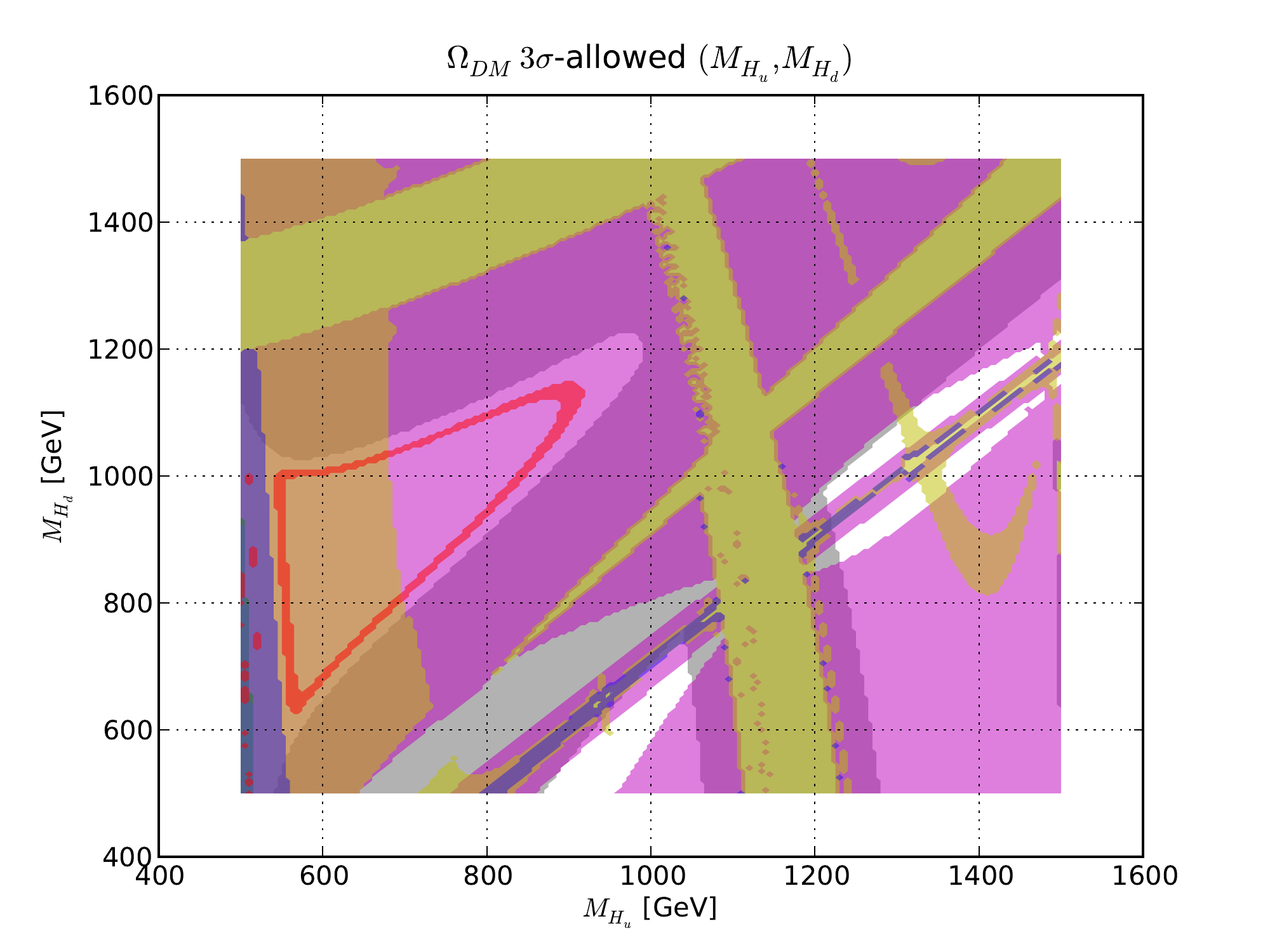}
\end{center}
\caption{Same as Figure~\ref{fig:dmscan}, except with $m_0=0$ for the third generation scalars as well.}
\label{fig:dmscan_light3rd}
\end{figure}

Second, the regions giving the correct relic density are far from those excluded by bounds from $b\rightarrow s\gamma$ and Higgs mass.  The former is not surprising since we have chosen small $\tan\beta$, $\mu>0$ and a gaugino soft mass that is not too light, but heavy $\tilde{t}$ and $\tilde{b}$ also suppress the relevant diagrams.  If $\tilde{t}$ and $\tilde{b}$ were nearly as light as the squarks of the first and second generation, the $b\rightarrow s\gamma$ limit would be exceeded for nearly all of the $(M_{H_u},M_{H_d})$ plane.  Similarly, the Higgs mass bound is also avoided rather easily because $\tilde{t}$ is heavy, enhancing the contribution of the $y_t$ term in the MSSM running of the $h_0$ mass, although the relationship to $(M_{H_u},M_{H_d})$ is more intricate.  For comparison, see Figure~\ref{fig:dmscan_light3rd}, which has the same parameters as Figure~\ref{fig:dmscan} except that $m_0=0$ for the third generation scalars as well.

Finally, we note that at small $m_0$ a gaugino soft mass $m_{1/2}\simeq 300$ has some phenomenological support, as it gives $\tilde\chi^0_{1,2}$ and $\tilde\chi_{1,2}^\pm$ masses favored by muon $g-2$~\cite{EO}.  Also, although a small $m_0$ usually gives dangerously small $\tilde \tau$ masses~\cite{EO}, this is not the case if it is only for the first and second generations.  This accounts for the difference between the gray regions of Figures~\ref{fig:dmscan}~and~\ref{fig:dmscan_light3rd}.

Altogether, these properties show that for the choices $\tan\beta=10$, $\mu>0$ and $A_0=0$, the NG hypothesis meets the relevant SUSY phenomenological requirements rather generically: several different regions of $(M_{H_u}, M_{H_d})$ at the correct mass scale can give the required relic density, and only a small portion of these regions is subject to other phenomenological constraints.  This is consistent with previous work which found that SUSY phenomenological requirements can be satisfied by mSUGRA with a sufficiently heavy third generation, albeit with equal $M_{H_u}$ and $M_{H_d}$~\cite{Baer}.  

We now consider how this scenario can be distinguished from conventional mSUGRA by examining the spectra of the benchmark points, whose $(M_{H_u}, M_{H_d})$ are shown in Table~\ref{table:bp}.  
\begin{table}
\begin{center}
\begin{tabular}{|c|c|c|}
\hline
Benchmark point & $M_{H_u}$ & $M_{H_d}$ \\
\hline
BP~1 & 860 & 1140 \\
BP~2 & 680 & 760 \\
BP~3 & 1115 & 1150 \\
\hline
\end{tabular}
\end{center}
\caption{Higgs soft masses for benchmark points, in GeV.}
\label{table:bp}
\end{table}

First, we choose a conventional mSUGRA point with similar soft masses for comparison:
\begin{equation}
m_0=75\;\rm{GeV},\quad m_{1/2} = 300\;\rm{GeV},\quad \tan\beta=10,\quad \mu > 0,\quad A_0=-200\;\rm{GeV}\;.
\end{equation}
This point, denoted CP, gives roughly the correct $\tilde\chi^0_1$ relic density through $\tilde\tau$ coannihilation and exhibits a small $\tilde l_R$--$\tilde \chi^0_1$ mass splitting.  The choice of $A_0=-200\;\rm{GeV}$ is to avoid Higgs mass bounds.  The resulting spectrum is shown pictorially in Figure~\ref{fig:spectr0}.  (A complete numerical listing of masses is given in Table~\ref{table:masses} of the Appendix, which starts on \pageref{appendix}.)
\begin{figure}
\begin{center}
\includegraphics[width=0.9\columnwidth]{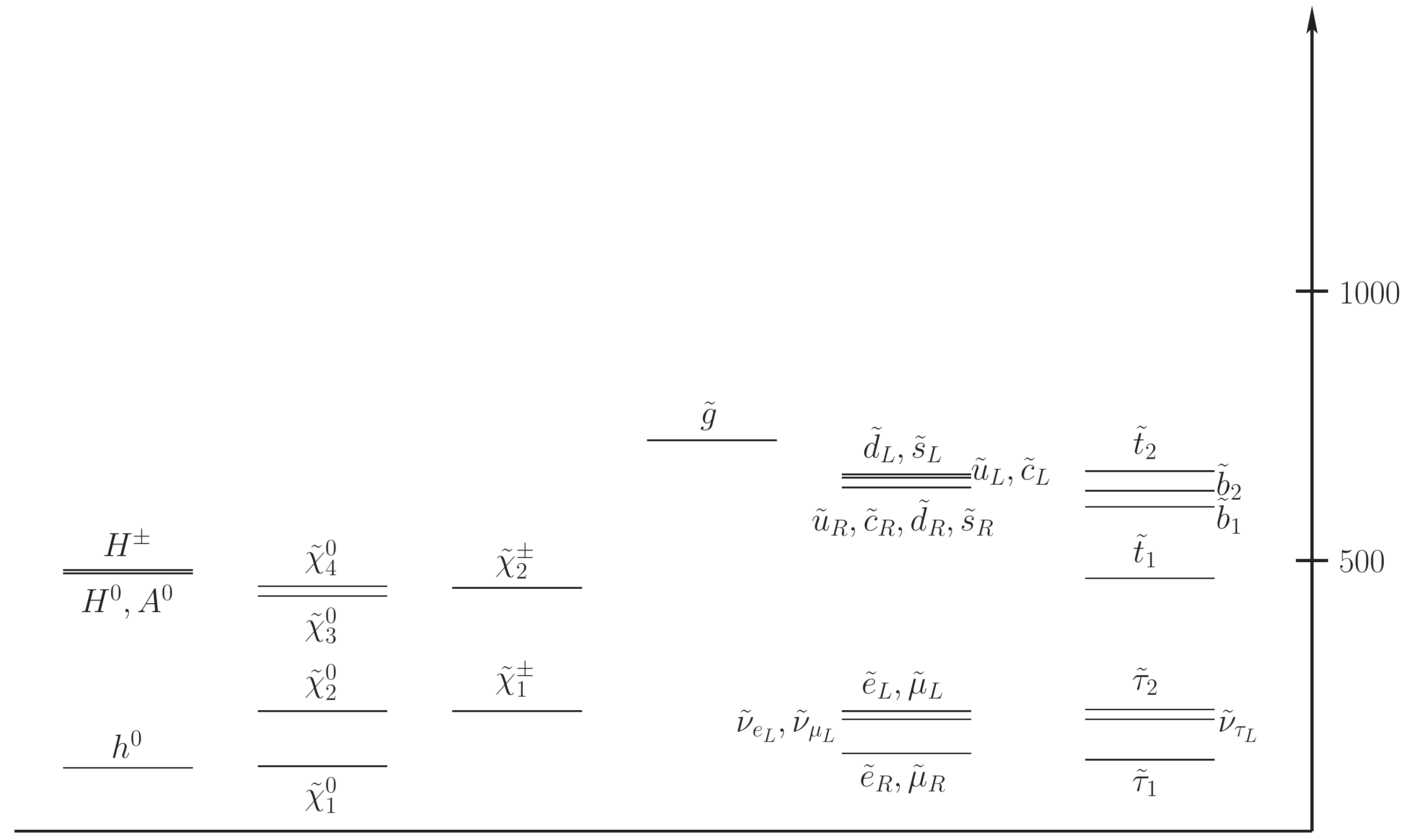}
\end{center}
\caption{Spectrum for comparison point (CP), in GeV.}
\label{fig:spectr0}
\end{figure}

\begin{figure}
\begin{center}
\includegraphics[width=0.9\columnwidth]{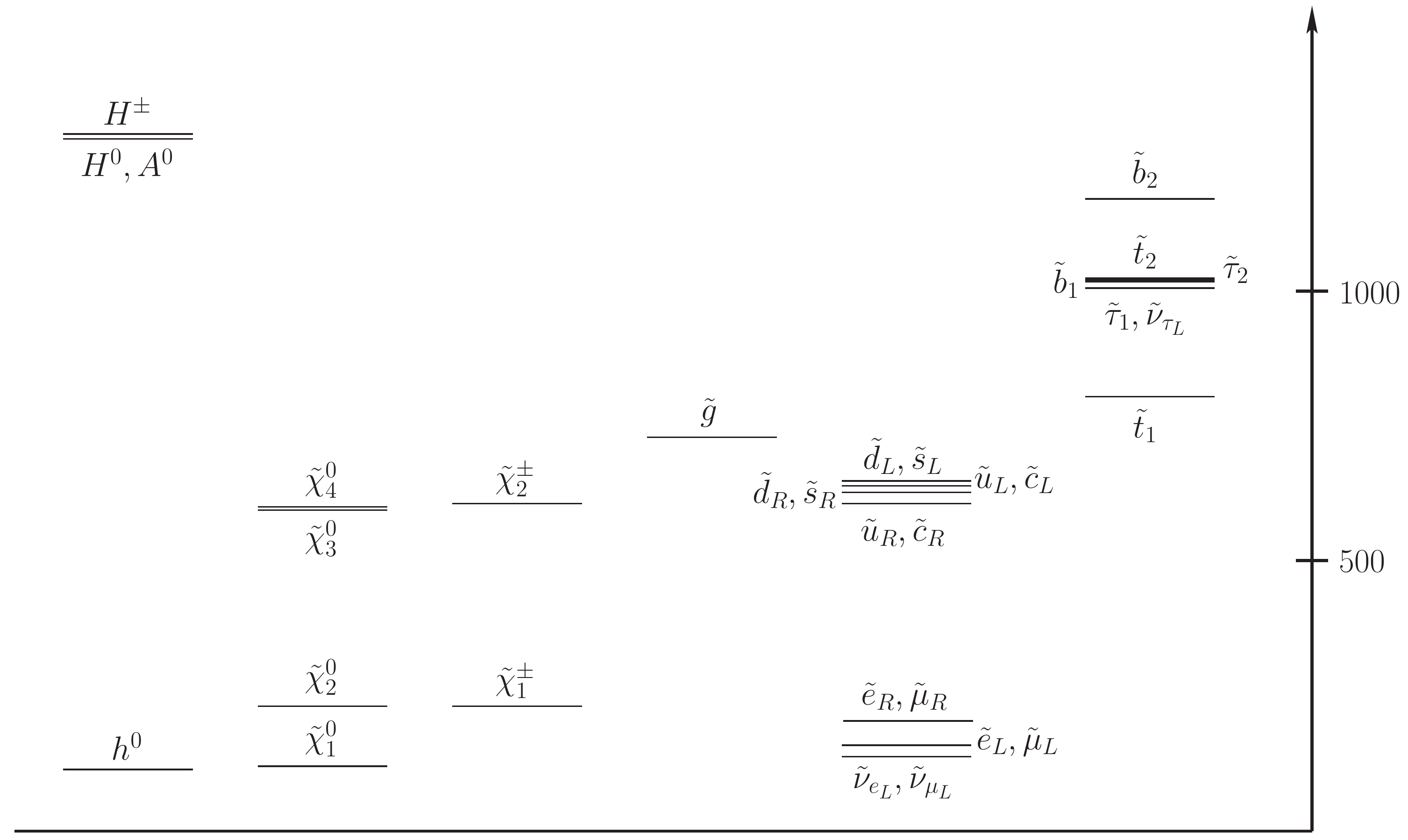}
\end{center}
\caption{Spectrum for benchmark point 1 (BP 1), in GeV.}
\label{fig:spect1}
\end{figure}

Let us consider benchmark point 1 (BP 1) shown in Figure~\ref{fig:spect1}.  As expected, compared to CP, the third generation scalars are very heavy, as are the heavy Higgs bosons due to the large values of $M_{H_u}$ and $M_{H_d}$.  
Also, like CP, there is a small splitting between the lightest slepton and $\tilde\chi^0_1$.  However, the expected hierarchy of sleptons is ``inverted,'' i.e. $\tilde\nu_{e,\mu}$ and $\tilde e_L$ are lighter than $\tilde l_R$.  This is due to the $S$ term in the running of the scalar masses (see, for example, Ref.~\cite{Martin}),
\begin{equation}
16\pi^2 \frac{d}{dt} m_{\phi_i}^2=-\sum_a 8 C_a(i)g_a^2\left|M_a\right|^2+\frac{6}{5}Y_i g_1^2 S
\end{equation}
where
\begin{equation}
\label{eq:S}
S\equiv {\rm Tr}\left[Y_j m_{\phi_j}^2\right ]=M_{H_u}^2-M_{H_d}^2+{\rm Tr}\left [{\bf m_Q^2}-{\bf m_L^2}-2{\bf m_{\bar u}^2}+{\bf m_{\bar d}^2}+{\bf m_{\bar e}^2}\right ]\;.
\end{equation}
In (\ref{eq:S}), the rightmost trace is zero for our case since $m_0$ is universal across all the scalars in a given generation, so $S=(M_{H_u}^2-M_{H_d}^2)$.  Since $\tilde l^*_R$ has hypercharge of $Y=+1$, but $\tilde l_L$ and $\tilde\nu_{e,\mu}$ have hypercharge of $Y=-1/2$, $\tilde l_R$ is driven heavier.  For the same reason, the spectrum of the left and right squarks of the first and second generations is brought closer together than in CP.  Unfortunately, the size of this effect is difficult to estimate analytically since $M_{H_u}$ and $M_{H_d}$ are themselves subject to significant running due to $S$.

\begin{figure}
\begin{center}
\includegraphics[width=0.9\columnwidth]{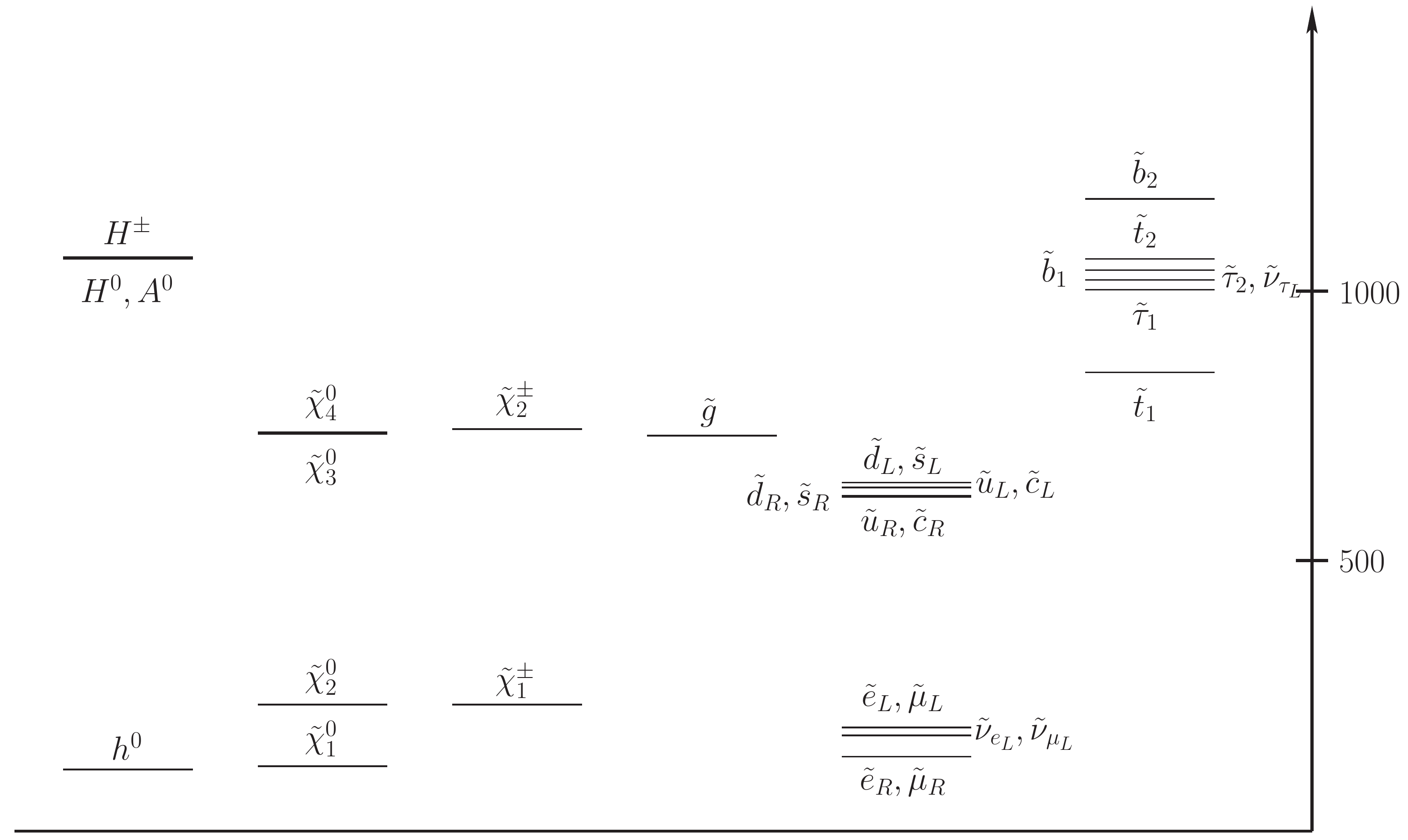}
\end{center}
\caption{Spectrum for benchmark point 2 (BP~2), in GeV.}
\label{fig:spect2}
\end{figure}

\begin{figure}
\begin{center}
\includegraphics[width=0.9\columnwidth]{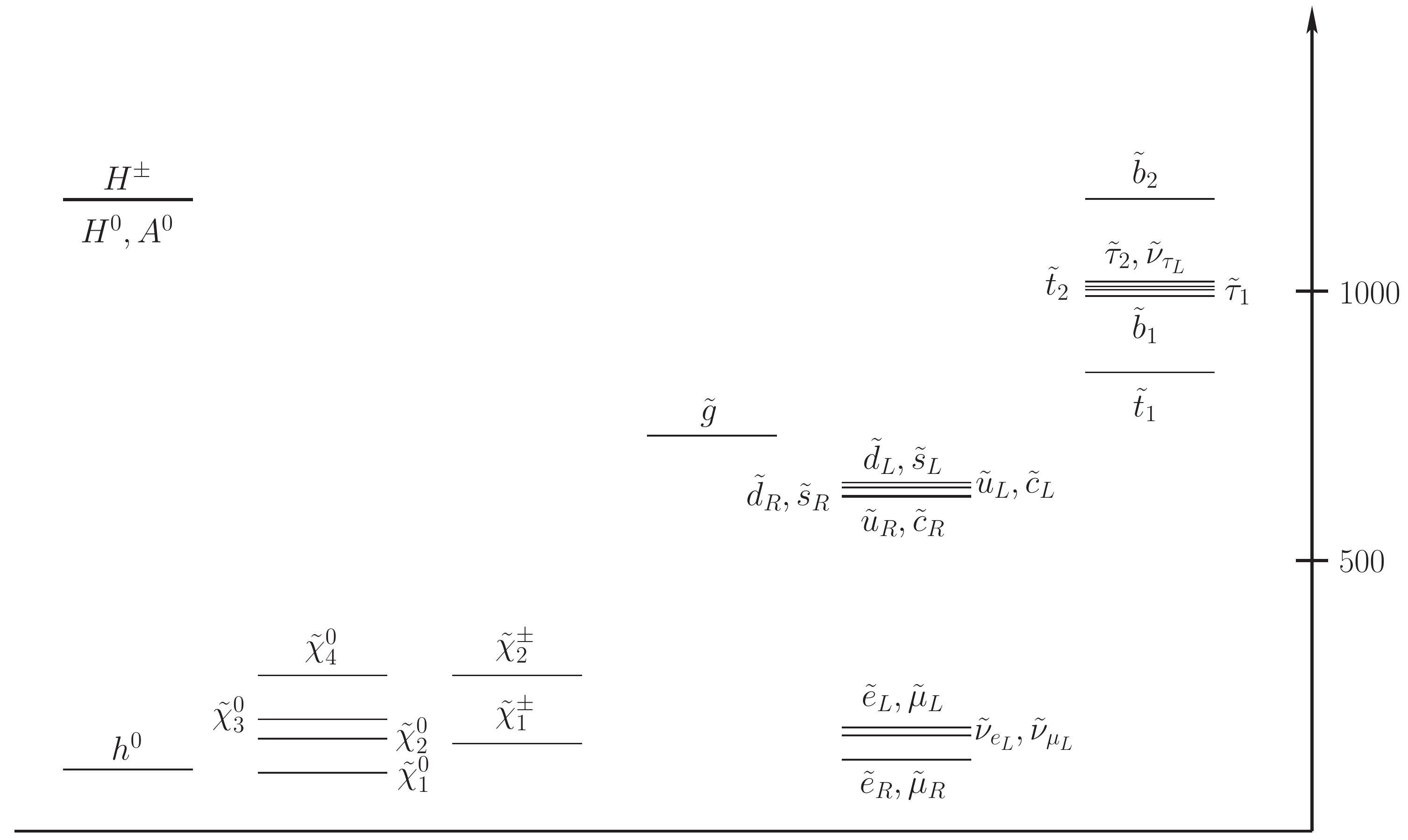}
\end{center}
\caption{Spectrum for benchmark point 3 (BP~3), in GeV.}
\label{fig:spect3}
\end{figure}

Benchmark point 2 (BP 2) in Figure~\ref{fig:spect2} is similar to BP~1, except that $(M_{H_u}^2-M_{H_d}^2)$ is smaller, so the slepton hierarchy is narrowed but remains non-inverted.
Benchmark point 3 (BP 3) in Figure~\ref{fig:spect3} is also similar, with $(M_{H_u}^2-M_{H_d}^2)$ being even smaller.  Except, in addition to a non-inverted and narrowed slepton hierarchy, because it is in the bridge branch it exhibits typical small-$\mu$ neutralino and chargino hierarchies.  In fact, $\tilde\chi_2^0$ and $\tilde\chi_1^\pm$ are lighter than $\tilde\nu_{e,\mu}$, closing off that decay channel.

In the next section, we discuss how these points can be identified at a collider, and use simulated events to demonstrate that this can be done early in 14~TeV running at the LHC.

\section{LHC physics}
\subsection{Discrimination at the LHC}
We now consider the distinguishing characteristics of NG hypothesis model points and their prospects for detection at the LHC.

Under the NG hypothesis, $m_3$ (i.e., $m_0$ for the third generation) is of a much higher order than $m_0$ for the first and second generations.  Therefore, the third generation scalars are significantly heavier than the squarks of the first and second generations.  On the other hand, the relationship between $m_{1/2}$ and $m_3$ is not established by the model.  While we have chosen $m_3=1$~TeV, the value could be much smaller, so the mass ordering between $\tilde g$ and the third generation scalars is not fixed. Consequently, the multiplicity of $b$-jets from gluino production is sensitive to the value of $m_3$.  

To study the effect of this uncertainty, in Table~\ref{table:gluinodecay} we list the branching ratios (\%) of gluinos decaying into modes with $b$-jets for $m_{1/2}=$ 300, 400 and 500~GeV.  For $m_3=0.7\,m_{1/2}$ the branching ratio is around 30\% because the decays are fully open.  However, if $m_3=1.3\,m_{1/2}$, the branching ratio is less than 1/4 of the $m_3=0.7\,m_{1/2}$ case.  At our choice of $m_{1/2}=300$~GeV and $m_3=1$~TeV the branching ratio is only ${\cal O}(1\%)$.
\begin{table}
\begin{center}
\begin{tabular}{|c||c|c|c|}
\hline
  $m_{1/2}$ GeV&   $m_3=m_{1/2}\times 0.7$ & $\times 1$  & $\times  
1.3$  \cr
\hline
  300 & 27.9 & 13.0 & 1.8 \cr
  400 & 31.5 & 18.2 & 3.2  \cr
500 &  33.5 & 20.9& 7.8 \cr
\hline
\end{tabular}
\caption{Branching ratios (\%) for $\tilde g$ decaying into $b$-jets for various $m_{1/2}$ and $m_3$.}
\end{center}
\label{table:gluinodecay}
\end{table}
By contrast, because the gaugino mass correction is smaller for slepton masses, the scalar mass of the order of gaugino mass at GUT scale is enough to make the third generation slepton heavier than the second lightest neutralino.  Therefore, $\tilde\tau$ would not be produced in neutralino and chargino decays irrespective of the value of $m_3$.  Consequently, few $\tau$-jets would be observed at the LHC, regardless of the relationsip between the $m_{1/2}$ and $m_3$.

Turning to leptons, because $\tilde\chi^0_2\rightarrow \tilde\tau \tau$ is closed, there will be many more events with the number of leptons $n_l \geq 2$ than for conventional mSUGRA points.  In the bridge branch, $\tilde\chi^0_2\rightarrow \tilde \nu_{e,\mu} \nu_{e,\mu}$ is closed as well, and $\tilde q\rightarrow q\tilde\chi_4^0$ and $\tilde q_L\rightarrow q_L\tilde\chi_2^\pm$ are open due to gaugino mixing caused by small $\mu$; however, most importantly, the right squarks have a large branching ratio to $\tilde\chi_2^0$ due to bino mixing.  Thus for BP~3 the number of $n_l \geq 2$ events is greatly enhanced, even over BP~1 and BP~2.  Overall, NG hypothesis model points require less integrated luminosity than conventional points in order to reconstruct two-lepton decay chains.  

In short, $\tau$-jet and lepton multiplicities can broadly distinguish NG hypothesis model points from conventional model points.  Moreover, a lack of $b$-jets would support the hypothesis, but a normal multiplicity of $b$-jets would not falsify it.

For greater precision, the effects of the $S$-term on the slepton mass hierarchy should be visible at the LHC.  At BP~1 and BP~2, the left squarks will decay mostly to $q_L\tilde\chi_2^0$ and $q_L\tilde\chi_1^\pm$ because these gauginos are mostly wino, and these in turn will decay mostly into left sleptons.  The right squarks will decay mostly to $q_R\tilde\chi_1^0$ because $\tilde\chi_1^0$ is mostly bino, so no leptons will be produced.  Thus, at these points we expect the leptons produced in the squark decay chain $\tilde q\rightarrow \chi^0_2 q \rightarrow q \tilde{l}^\pm l^\mp$ to be mostly left-handed.  By comparison, at CP $\tilde\chi_2^0\rightarrow \tilde l_R^\pm l^\mp$ exclusively since $\tilde l_L$ is too heavy --- most of these leptons will be right-handed.

Because the chirality of the lepton is fixed by the interaction vertex, the direction of the lepton in the rest frame of the squark is  correlated with the quark and lepton chirality.  Namely, for the final states $q_L l_L$ ($q_R l_R$, $q_L \bar{l}_R$, $q_R \bar{l}_L$) the jet and lepton tend to go to in opposite directions while for the $q_L l_R$ ($q_R l_L$, $q_L \bar{l}_L$, $q_R \bar{l}_R$) final state they tend go in the same direction.  Since the LHC is a $p$--$p$ collider, more $\tilde q$ is produced than $\tilde q^*$, so there will be a charge  
asymmetry in the distribution of the jet--lepton invariant mass.  The lepton from the slepton decay is not correlated since the slepton is scalar.  Then, for BP~1 and BP~2 , there will be a large charge asymmetry since most of the $\tilde\chi^0_2$ are from $\tilde q_L$ rather than $\tilde q_R$, and $\tilde\chi^0_2$ decays  predominantly to $\tilde{l}_Ll_L$.  Note that the charge asymmetry is opposite for CP because here the dominant channel is $\tilde q_L\rightarrow \tilde\chi^0_2\rightarrow \tilde l_R$.

The situation is more complicated at BP~3 because so many different squark decay channels are open.  Moreover, because $\tilde\chi_2^0$ is so light, it can only decay to $\tilde l_R$.  We first observe that the right squarks produce leptons mostly through $\tilde\chi_2^0$ (22\%) with a lower branching fraction than left squarks (17\% for up, 10\% for down). The charge asymmetry from right squarks is that of $\tilde q_R\rightarrow \tilde\chi^0_2\rightarrow \tilde l_R$, the same sign as BP~1 and BP~2, but not as large.  To clearly distinguish BP~3 from BP~1 and BP~2, its small-$\mu$ character can be seen by observing the tight neutralino mass hierarchy through the small $\tilde\chi_2^0$--$\tilde\chi_1^0$ mass splitting.  In addition, the decay of squarks to heavier neutralinos can be observed.  For example, $\tilde q_L$ decay into $\tilde\chi_4^0$ and $\tilde\chi^\pm_{1,2}$ with branching ratios 15\% and 65\%, respectively.%\footnote{Notable ranching ratios for all points are shown in Tables~\ref{table:bp1br}, \ref{table:bp2br}, \ref{table:bp3br},\ref{table:cpbr} of the Appendix.}

Lastly, at all the benchmark points the heavy states in the Higgs sector will not be produced in sufficient quantities to be identified because their mass is too high.  However, if $\tan\beta > 15$, a heavy Higgs of ${\cal O}(500\;{\rm GeV})$ may be observed at a CP-like model point~\cite{ATLASTDR}.

\subsection{Simulation and reconstruction}
To demonstrate this phenomenology at the LHC, for each of the model points we generated $10^5$ signal events at 14~TeV for inclusive squark production using the {\tt ISAJET} (version 7.72) spectrum calculator~\cite{ISAJET}, {\tt HERWIG} (version 6.5) shower generator~\cite{HERWIG} and {\tt AcerDET} (version 1.0)~\cite{AcerDET} fast detector simulation.  (MSSM input parameters were tuned slightly such that {\tt ISAJET} would produce the same spectrum as {\tt SOFTSUSY}.)  The inclusive production cross section is $\gsim 15\;{\rm pb}$, so the integrated luminosity required for the following analysis is only $7\;{\rm fb}^{-1}$; this should be achievable in roughly one month of running~\cite{LHCDR}.

To grossly distinguish NG hypothesis model points from conventional points we first extract and compare:
\begin{itemize}
\item number of $b$-jets $n_b$ with $p_T > 50\;\rm{GeV}$
\item number of $\tau$-jets $n_\tau$ with $p_T > 20\;\rm{GeV}$
\item number of leptons $n_l$ with $p_T > 15\;\rm{GeV}$ and $\eta < 2$ .
\end{itemize}
Here, the $b$- and $\tau$-jets are the jets labeled as such\footnote{Only hadronic $\tau$ decays passing certain cuts are labeled as $\tau$-jets by {\tt AcerDET}.  Thus, leptonic $\tau$ decays are not included in $n_\tau$.  See the {\tt AcerDET} manual~\cite{AcerDET} for details.} by {\tt AcerDET}, with an efficiency of around 80\%.  However, real tagging efficiencies are probably around 60\% and 50\% for $b$- and $\tau$-jets, respectively~\cite{AcerDET}.  
The results are shown in Table~\ref{table:nprods}, as a percentage of events.
\begin{table}
\begin{center}
\begin{tabular}{|c|r|r|r|}
\hline
Model point & \% $n_b\geq 1$ & \% $n_\tau\geq 1$ & \% $n_l\geq 2$\\
\hline
BP~1 & 2.29 & 0.04 & 15.80 \\
BP~2 & 2.22 & 0.02 & 15.84 \\
BP~3 & 3.17 & 2.284 & 28.97 \\
\hline
CP & 31.70 & 16.94 & 4.10 \\
\hline
\end{tabular}
\end{center}
\caption{Percentage of events with the given multiplicities (see text for details).}
\label{table:nprods}
\end{table}
As expected, for the benchmark points there are a few $b$-jets and very few $\tau$-jets from direct production of third generation squarks.  BP~3 shows an enhancement in $\tau$-jets from off-shell $\tilde\chi_1^\pm$ decays via the $W$ boson.  By contrast, CP has many $b$-jets from both gluino and third generation squark decays, as well as many $\tau$-jets (and therefore fewer multi-leptons) from $\tilde\chi_2^0$ decays.

Turning to the other discriminators, in order to obtain the $\tilde\chi$--$\tilde l$ mass splittings and charge asymmetry, we must study the gluino and squark decay chains
\begin{equation}
(\tilde g\rightarrow)\;\tilde q(j)\rightarrow \tilde\chi_2^0jj\rightarrow \tilde l^\pm jjl_1^\mp\rightarrow jjl_1^\mp l_2^\pm+\not\!\!E_T\;.
\end{equation}
Henceforth, we define the lepton $l_1^\mp$ from $\tilde\chi_2^0$ decay as the ``near'' lepton, and the lepton $l_2^\pm$ from $\tilde l$ decay to be the ``far'' lepton.
We use the method of invariant mass distributions and endpoints to reconstruct the masses~\cite{ATLASTDR,endpoints,GMO}.  Here, one creates distributions of the invariant masses $m_{p_1p_2\dots}$ of the final state particles, then identifies the upper endpoint of each distribution $m_{p_1p_2\dots}^{max}$.  These endpoint values are then inverted to obtain the masses of the particles in the decay chain.  We use the following invariant masses in our reconstruction:
\begin{itemize}
\item $m_{ll}$, the invariant mass of both leptons
\item $m_{jll}$, the invariant mass of both leptons $+$ the jet from squark decay
\item $m_{jl(lo)}\equiv \min (m_{jl_1},m_{jl_2})$, where each $m_{jl_i}$ is the invariant mass of the jet from squark decay and one of the leptons
\item $m_{jl(hi)}\equiv \max (m_{jl_1},m_{jl_2})$.
\end{itemize}
This method can determine the $\tilde l$--$\tilde\chi_1^0$ and $\tilde\chi_2^0$--$\tilde l$ mass splittings rather precisely, which is useful here; extracting the squark mass scale is less critical because it is mostly determined by $m_0$.  We use the {\tt MINUIT2} fitter in {\tt ROOT}~\cite{ROOT} to find the endpoints, and the inversion formulas in Ref.~\cite{GMO} to find the mass differences and squark mass.  

For each model point we identify events with $n_l\geq 2,\; n_j\geq 2$ where the two highest $p_T$ leptons are same-flavor/opposite-sign (SFOS).  To subtract the chargino contribution to SFOS, we also identify $n_l=2$ events with different-flavor/opposite-sign (DFOS) and subtract these counts from the SFOS invariant mass distributions before fitting.  In a given event, we choose the jet among the two highest $p_T$ jets that gives the smallest $m_{jll}$.  This identifies the jet from squark decay on the correct branch.

\begin{figure}
\begin{center}
\includegraphics[width=0.45\columnwidth]{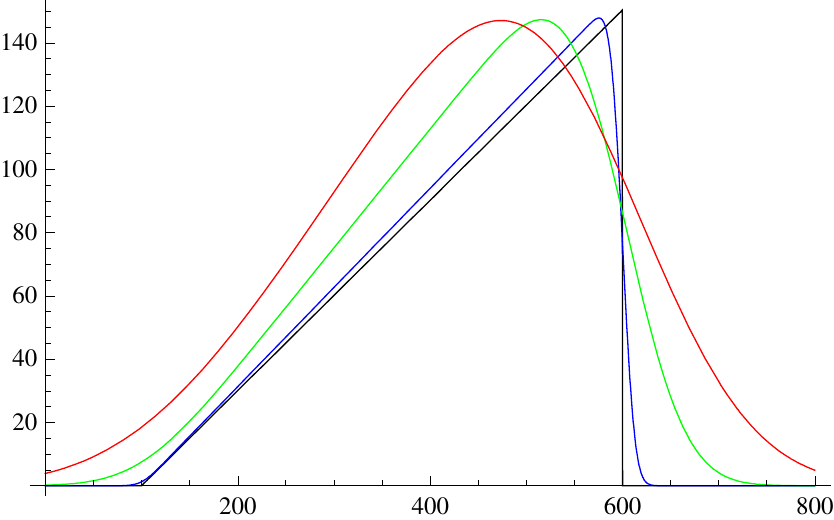}
\includegraphics[width=0.45\columnwidth]{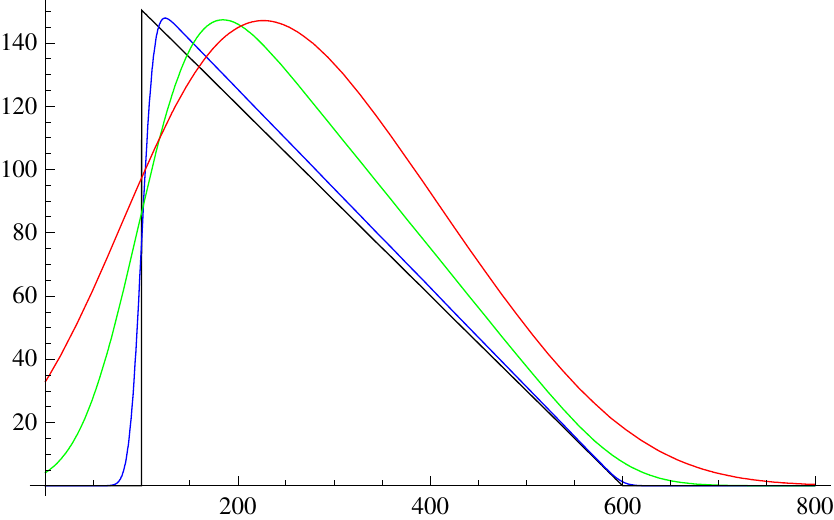}
\end{center}
\caption{Example fitting functions with endpoint = 600 and gaussian smearing widths = 0 (black), 10 (blue), 50 (green) and 100 (red).  On the left the endpoint shape is a vertical edge, whereas on the right it is a ramp.}
\label{fig:smeared_ramp}
\end{figure}

Noting the example mass distribution shapes in Ref.~\cite{GMO}, we use the vertical edge with gaussian smearing in order to fit the $m_{ll}^{max}$ endpoint, and a ramp function with gaussian smearing in order to fit the $m_{jll}^{max}$ and $m_{jl(lo)}^{max}$ endpoints (for examples showing both kinds of shapes, see Figure~\ref{fig:smeared_ramp}). 
We also choose a vertical edge for $m_{jl(hi)}$ because $m_{jl(near)}$ (the jet--near lepton invariant mass distribution) has edges.  This is a good choice for the BP~1, BP~3 and CP because $m_{jl(near)}^{max}>m_{jl(far)}^{max}$.  At point BP~2, $m_{jl(near)}^{max}<m_{jl(far)}^{max}$, so this endpoint shape does not provide a good fit.  In this case, the $m_{jl(hi)}$ distribution is much closer in shape to a ramp function near the end point, and it has a discontinuity at $m_{jl(hi)}\sim m_{jl(lo)}^{max}$.  We call this discontinuity a ``secondary edge.''  We try both the smeared ramp and a smeared vertical edge to fit this endpoint.

In our fits the smearing width is left as a free parameter to match the broadening of the endpoint due to energy uncertainty and other effects included in the simulation which reduce the number of events near the endpoint; this is most apparent in the $m_{jl(hi)}^{max}$ fits.  

In addition, for BP~3, there are significant tails in the $m_{jl(hi)}$ and $m_{jll}$ distributions due to the decays of $\tilde\chi_4^0$ and $\tilde\chi_2^\pm$.  In these fits the fitting function is a smeared ramp added to a sloping line in piecewise fashion.  Moreover, in these functions we set the maximum smearing width to 5~GeV and over a limited mass range to avoid over-fitting to the curves of the edges which are resolvable due to the large statistics.

\begin{figure}
\begin{center}
\includegraphics[width=0.7\columnwidth]{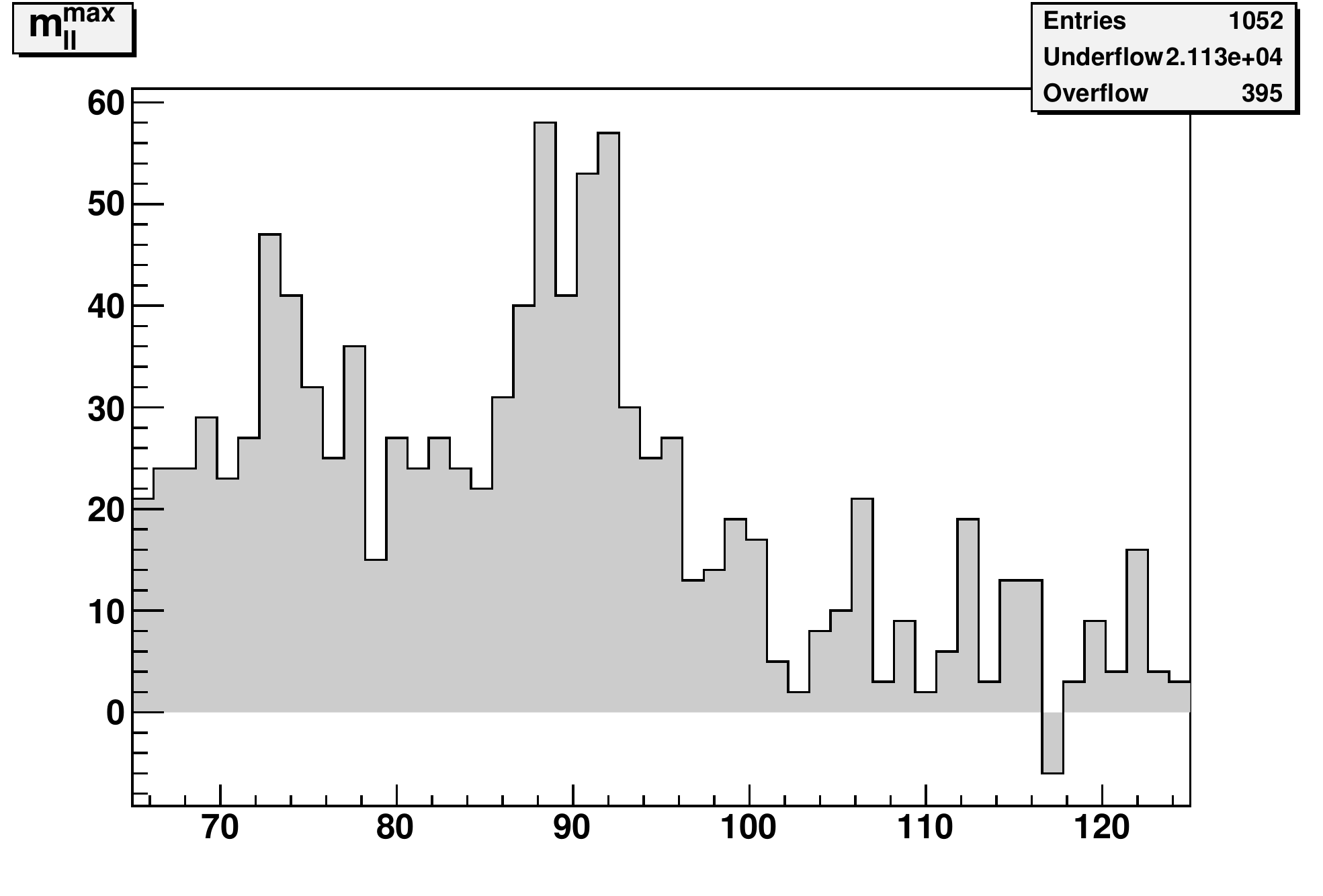}
\end{center}
\caption{Higher range of $m_{ll}$ for BP~3.  At 93~GeV is the $\tilde\chi_2^\pm\rightarrow\tilde\nu_L l^\pm\rightarrow\tilde\chi_1^\pm ll$ endpoint, and at 107~GeV is the $\tilde\chi_4^0\rightarrow\tilde l_L l\rightarrow\tilde\chi_2^0 ll $ endpoint.}
\label{fig:pt3_higher_endpoints}
\end{figure}

The $\tilde\chi_4^0$ and $\tilde\chi_2^\pm$ edges of BP~3 can be seen in the $m_{ll}$ mass distribution in Figure~\ref{fig:pt3_higher_endpoints}.  At 93~GeV is the $\tilde\chi_2^\pm\rightarrow\tilde\nu_L l^\pm\rightarrow\tilde\chi_1^\pm ll$ endpoint, and at 107~GeV is the $\tilde\chi_4^0\rightarrow\tilde l_L l\rightarrow\tilde\chi_2^0 ll $ endpoint.
All the other plots of the mass distributions and their fit endpoints, as well as all the inversion solutions, are shown in the Appendix %(Figures~\ref{fig:bp1}, \ref{fig:bp2}, \ref{fig:bp3}, \ref{fig:cp}; Tables~\ref{table:bp1ep}, \ref{table:bp1ms}, \ref{table:bp2ep}, \ref{table:bp2ms}, \ref{table:bp3ep}, \ref{table:bp3ep2}, \ref{table:bp3ms}, \ref{table:cpep}, \ref{table:cpms})
. In these plots the cross-hatched region shows the parton-level mass distributions, and the solid region the jet-level distributions.  One observes that the fitting function identifies the $m_{ll}^{max}$ endpoints correctly without any shifts.  Also, one can see that the jet-level distributions have fewer counts in the bins near the endpoint and more counts in the lower energy bins.  This is due to effects such as initial state radiation and jet misidentification.

 Here, we show the correct solutions in Table~\ref{table:mdiffs}, where the values in the parentheses are the true values.
\begin{table}
\begin{center}
\begin{tabular}{|c|c|c|c|c|}
\hline
Model point & $m_{\tilde l}-m_{\tilde\chi_1^0}$ & $m_{\tilde\chi_2^0}-m_{\tilde l}$ & $m_{\tilde q}-m_{\tilde\chi_2^0}$ & $m_{\tilde q}$ \\
\hline\hline
BP~1 & $37.7\pm 1.1$ (37.9) & $73.3\pm 1.9$ (73.5) & $433\pm 7$ (428) & $655\pm 35$ (660) \\ \hline
BP~2 & $72.9\pm 0.7$ (73.5) & $40.8\pm 0.3$ (40.8) & $414\pm 3$ (409) & $660\pm 10$ (643) \\ \hline
BP~3 & $29.1\pm 1.0$ (28.9) & $31.4\pm 0.9$ (31.6) & $491\pm 12$ (472) & $693\pm 88$ (639) \\
\hline\hline
CP & $27.7\pm 3.3$ (25.7) & $76.0\pm 4.5$ (77.5) & $497\pm 51$ (444) & $848\pm 237$ (666) \\
\hline
\end{tabular}
\end{center}
\caption{Reconstructed mass differences and squark masses, in GeV.  True values are shown in parentheses.}
\label{table:mdiffs}
\end{table}
The correct solution is chosen by first noting that none of the model points have off-shell decays in the squark decay chains, which would manifest in flat mass distributions.  This eliminates Region~4 of $m_{jll}^{max}$, in the terminology of Ref.~\cite{GMO}.  Next, among the Region~1 solutions we choose the one with the plausible squark mass that would give the observed production cross-section.  If this is ambiguous, looking at the $m_{jl(hi)}$ mass distribution, if there is a secondary edge then the lepton at the fit edge is the far lepton, and so the $\tilde l$--$\tilde\chi_1^0$ splitting is larger than the $\tilde\chi_2^0$--$\tilde l$ splitting; conversely, if there is no secondary edge, the fit lepton is the near lepton, and the $\tilde\chi_2^0$--$\tilde l$ splitting is the larger one.

Finally, to calculate the charge asymmetry, we create $m_{jl^\pm}$ distributions separately for both of leptons, where $j$ is the same jet as from $m_{jll}$.  This gives us four distributions $m_{jl^\pm(soft)}$ and $m_{jl^\pm(hard)}$ where the former distribution is for the lepton with the lower $p_T$ of the two, and the latter is for the lepton with the greater $p_T$ (these are also shown in the Appendix).  Then, we integrate each of the distributions,
\begin{equation}
N(l)=\int_a^b dm'\;m_{jl}(m')
\end{equation}
where $(a,b)=(m_{jl(lo)}^{max}-100\;{\rm GeV}, m_{jl(lo)}^{max})$ for the soft distribution.  For the hard distribution, $(a,b)=(m_{jl(near)}-100\;{\rm GeV}, m_{jl(near)})$ where $m_{near}$ is the edge of the $m_{jl(hi)}$ distribution for the near lepton.  For our model points, $m_{jl(near)}\neq m_{jl(hi)}^{max}$ only for BP~2, which has a secondary edge at 350~GeV.  

This gives us four quantities $N_{soft}(l^\pm)$ and $N_{hard}(l^\pm)$.  Finally, we define the charge asymmetry\footnote{Our definition has the opposite sign from some others in the literature.} as
\begin{equation}
A_c\equiv \frac{N(l^-)-N(l^+)}{N(l^-)+N(l^+)}\;.
\label{eq:chargeasymm}
\end{equation}
The soft and hard lepton charge asymmetries, $A_{c,soft}$ and $A_{c,hard}$, are shown for each of the model points in Table~\ref{table:chargeasymm}.
\begin{table}
\begin{center}
\begin{tabular}{|c|r|r|}
\hline
Model point & $A_{c,soft}\quad$ & $A_{c,hard}\quad$ \\
\hline
BP~1 & $-0.11\pm 0.02$ & $0.25\pm 0.02$ \\
BP~2 & $-0.04\pm 0.02$ & $0.11\pm 0.02$ \\
BP~3 & $-0.01\pm 0.01$ & $0.04\pm 0.01$ \\
\hline
CP & $0.04\pm 0.04$ & $-0.14 \pm 0.06$ \\
\hline
\end{tabular}
\end{center}
\caption{Lepton charge asymmetries for each model point.}
\label{table:chargeasymm}
\end{table}
We see that $\left|A_{c,soft}\right|<\left|A_{c,hard}\right|$, since more of the soft leptons are from slepton decay.

\subsection{Model point identification}
Using the results from the reconstruction procedures above, we describe how to identify the different model points from the data.  We will assume universal gaugino soft masses at high scale, such that $M_3:M_2:M_1\approx 6:2:1$ at low scale~\cite{Martin}.
\subsubsection*{BP~1}
The single edge of the $m_{jl(hi)}$ distribution suggests that this is from the near lepton.  The large $\tilde\chi_2^0$--$\tilde\chi_1^0$ mass splitting and non-observation of higher $m_{ll}$ edges from $\tilde\chi_4^0$ and $\tilde\chi_2^\pm$ decays indicate that $\mu > M_2-M_1$, so $\tilde\chi_2^0$ is gaugino-dominated and the decay chains are primarily from left squarks.  Combined with the large positive  (by our definition, Eq.~\ref{eq:chargeasymm}) charge asymmetry, this suggests that the slepton for this edge is $\tilde l_L$.  Following the mSUGRA mass relations~\cite{Martin}, the large $\tilde\chi_2^0$--$\tilde l_L$ splitting is inconsistent with mSUGRA assumptions; on the other hand, the $\tilde l_L$--$\tilde\chi_1^0$ splitting is small enough to give the correct dark matter relic abundance.  Then, we can infer that $\tilde l_R$ is heavier than $\tilde l_L$.  Along with the lack of $b$- and $\tau$-jets, an upper branch point in the NG hypothesis is a plausible candidate.
\subsubsection*{BP~2}
Again, the $\tilde\chi_2^0$--$\tilde\chi_1^0$ splitting is large and no higher $m_{ll}$ edges are observed, so we are not in a small-$\mu$ region and left squark decays produce the most leptons.  The secondary edge at 350~GeV in the $m_{jl(hi)}$ distribution suggests that the fitted edge is for the far lepton.  Combined with the large $\tilde l$--$\tilde\chi_1^0$ mass splitting and positive charge asymmetry, we infer that this edge is for the $\tilde l_L$ and that there is an unresolved $\tilde l_R$ close to $\tilde\chi_1^0$, giving the correct dark matter relic abundance.  The smaller charge asymmetry than for BP~1 is consistent with the far lepton being more likely to be the hard lepton, contributing roughly 50\% below 350~GeV in the $m_{jl(hi)}$ distribution.  Thus, in this case, a lower branch point in the NG hypothesis is a plausible candidate.
\subsubsection*{BP~3}
The additional edges in the $m_{ll}$ distribution due to the higher neutralinos and the small $\tilde\chi_2^0$--$\tilde\chi_1^0$ mass splitting suggest a small-$\mu$ point.   These edges can be reconstructed to obtain the masses of these heavy neutralinos~\cite{ChargeAsymm}.  Combining the inferred branching ratios, we can determine that right squark decays contribute more strongly to the the charge asymmetry.  Since the asymmetry is positive, this suggests that the reconstructed slepton is $\tilde l_R$; we can reconstruct $\tilde l_L$ and $\tilde\nu_L$ at some specific higher mass.  A bridge branch point is consistent with all these features, as well as with the enhancement in the number of multi-lepton events over BP~1 and BP~2.
\subsubsection*{CP}
Unlike the three scenarios above, here we observe a typical number of $b$- and $\tau$-jets along with few multi-lepton events.  Since the $\tilde\chi_2^0$--$\tilde\chi_1^0$ splitting is large, this is not a small-$\mu$ point, so most of the multi-lepton events are from left squark decays.  Combined with the negative charge asymmetry, we infer that the slepton for this edge is $\tilde l_R$.  We also see that the lepton is near, so it is the $\tilde\chi_2^0$--$\tilde l_R$ splitting which is large.  Then, $\tilde l_L$ has some mass greater than that of $\tilde \chi_2^0$, otherwise the charge asymmetry would be washed out or positive.  This is likely a typical mSUGRA point.

In summary, the benchmark points of the NG hypothesis from different regions in the $(M_{H_u},M_{H_d})$ plane can be readily identified in early 14~TeV running at the LHC, and are easily distinguished from typical mSUGRA points.

\section{Conclusions and outlook}
The hierarchy of the Yukawa couplings in the standard model remains an open question.  We have presented a class of models wherein the first and second generation fermions are SUSY partners of Nambu-Goldstone bosons which parameterize a non-compact K\"ahler manifold, such that the first and second generation Yukawa couplings are forbidden by the low-energy theorem. Then we gave an (incomplete) example to show that such a model can be constructed.  

Next, we found that many different model points in this scenario can give the correct dark matter abundance while easily evading phenomenological bounds, and examined the spectra of benchmark points in different regions of the allowed parameter space.  Finally, we argued that these points can be distinguished from conventional mSUGRA points at the LHC, and demonstrated this assuming only $7\;{\rm fb}^{-1}$ of integrated luminosity at 14~TeV, with $m_{\tilde g}\sim 700\;{\rm GeV}$.

Nonetheless, improvements can be made.  First, explicit models should be constructed, and the consequences of their gravitino and novino fields investigated.  Also, the running of the soft masses between the SUSY-breaking and GUT scales should be verified to be small.  Second, a more expansive simulation (including backgrounds) should be done with higher integrated luminosity to improve the quality of the endpoint fits and increase the significance of the charge asymmetries.  Third, observability at LHC should be verified across the entire $(M_{H_u}, M_{H_d})$ plane, not only for specific benchmark points.

Finally, we close with two notes on non-LHC phenomenology.  First, concerning signatures of dark matter, at model points BP~1 and BP~2 the LSP would be very difficult to detect outside of the LHC.  The $s$-wave annihilations of neutralinos to fermions are helicity-suppressed, and because the LSP at BP~1 and BP~2 is mostly bino, annihilations to the standard model weak bosons are also suppressed.  Thus, the total annihilation cross-section is only ${\cal O}(10^{-30})\;{\rm cm}^3\;{\rm s}^{-1}$ in the present universe, far beyond the reach of current or planned astrophysical observatories.  On the other hand, at BP~3, the large higgsino component gives a large annihilation cross-section ${\cal O}(10^{-26})\;{\rm cm}^3\;{\rm s}^{-1}$ to $W^+W^-$.  Combined with the low mass, the LSP of BP~3 may be able to explain the anomalous rise in the positron fraction observed by PAMELA~\cite{Adriani:2008zr,CirelliPK}; however, the bound from antiprotons~\cite{Adriani:2008zq} must also be checked.  Moreover, as discussed, the large higgsino component also enhances the direct detection cross-section.  The predicted cross-section $2\times 10^{-8}$~pb is near current limits and therefore within the reach of forthcoming experiments.  By contrast, for BP~1 and BP~2, the predicted cross-sections are only $1\times 10^{-10}$~pb and $2\times 10^{-10}$~pb, respectively.

Second, the flavor rotations in the quarks and leptons generate flavor-changing transitions among the squarks and sleptons.  However, the transition parameter $\delta_{12}$~\cite{Gab} between the first and the second generation is suppressed by $\lambda^5$, where $\lambda$ is the Wolfenstein parameter~\cite{Wolfe} in the rotation matrices. Taking $\lambda_d\simeq 0.1$ we see the prediction is very close to the present upper bound on $\delta_{12}$. Interestingly, the transition between the strange and bottom quark is suppressed only by $\lambda_d^2$, which may generate some sizable contributions to the $B_s-\bar B_s$ mixing matrix.

In conclusion, we have shown that the NG hypothesis may explain the smallness of the masses of the first and second generation standard model fermions, and that this hypothesis can be tested at the LHC.

\section*{Acknowledgments}
SKM and MS are supported by World Premier International Research Center Initiative (WPI Initiative), MEXT, Japan.  TTY would like to thank T.~Kugo for useful discussions.

\clearpage
\appendix
\section{Model point data}
\label{appendix}
In this section we give miscellaneous data on the model points:
\begin{itemize}
\item Complete listing of masses of new particles (for all model points)
\item Plots of invariant mass distributions, with edge fits
\item Expected and fit endpoints (as well as secondary endpoints)
\item All mass solutions
\item Notable branching ratios
\end{itemize}
{\bf Note:}  All mass units are in GeV.
\clearpage
\subsection{Complete listing of masses of new particles}
\begin{table}[h]
\begin{center}
\begin{tabular}{|c|c|c|c|c|}
\hline
{\bf Particle} & {\bf BP 1} & {\bf BP 2} & {\bf BP 3} & {\bf CP}\\

\hline
$\tilde d_L$, $\tilde s_L$ & 660.4 & 643.6 & 639.2 & 660.2\\
$\tilde u_L$, $\tilde c_L$ & 655.5 & 638.6 & 634.2 & 655.3\\
$\tilde b_1$ & 1026 & 1041 & 998 & 601.8\\
$\tilde t_1$ & 801.6 & 847.7 & 742.5 & 467.8\\
$\tilde l_L$ & 158.2 & 193.7 & 194.0 & 222.1\\
$\tilde {\nu_l}_L$ & 136.6 & 176.5 & 177.0 & 207.4\\
$\tilde \tau_1$ & 1005 & 1000 & 999.9 & 132.6\\
$\tilde {\nu_\tau}_L$ & 1006 & 1013 & 1012 & 206.1\\
\hline
$\tilde d_R$, $\tilde s_R$ & 641.7 & 620.4 & 616.5 & 634.9\\
$\tilde u_R$, $\tilde c_R$ & 619.5 & 616.8 & 613.5 & 635.6\\
$\tilde b_2$ & 1159 & 1159 & 1156 & 629.2\\
$\tilde t_2$ & 1045 & 1063 & 1014 & 665.8\\
$\tilde l_R$ & 205.3 & 138.3 & 135.3 & 144.5\\
$\tilde \tau_2$ & 1016 & 1020 & 1016 & 225.2\\
\hline
$\tilde g$ & 730.9 & 729.3 & 731.0 & 721.7\\
$\tilde \chi^0_1$ & 120.3 & 120.2 & 106.3 & 118.8\\
$\tilde \chi^0_2$ & 231.7 & 234.5 & 166.9 & 222.0\\
$\tilde \chi^\pm_1$ & 232.1 & 234.9 & 160.4 & 222.2\\
$\tilde \chi^0_3$ & 597.8 & 737.4 & 203.8 & 434.5\\
$\tilde \chi^0_4$ & 606.7 & 743.6 & 286.4 &  451.7\\
$\tilde \chi^\pm_2$ & 607.2 & 744.3 & 282.0 & 450.7\\
\hline
$h^0$ & 115.0 & 115.3 & 114.6 & 116.0\\
$H^0$ & 1293 & 1071 & 1176 & 476.1\\
$A^0$ & 1284 & 1063 & 1168 & 472.8\\
$H^\pm$ & 1295 & 1073 & 1179 & 482.5\\
\hline
\end{tabular}
\end{center}
\caption{Listing of masses for model benchmark points and mSUGRA comparison point.}
\label{table:masses}
\end{table}
\clearpage

\subsection{Benchmark point 1 (BP 1)}

\subsubsection{True masses}
$$m_{\tilde\chi_1^0}=120.3,\quad m_{\tilde l}=158.2,\quad m_{\tilde\chi_2^0}=231.7,\quad m_{\tilde q}=660.4$$
%\begin{table}[h]
%\begin{center}
%\begin{tabular}{|c|c|c|c|}
%\hline
%$m_{\tilde\chi_1^0}$ & $m_{\tilde l}$ & $m_{\tilde\chi_2^0}$ & $m_{\tilde q}$\\
%\hline
%120.3 & 158.2 & 231.7 & 660.4\\
%\hline
%\end{tabular}
%\end{center}
%\caption{\label{tab:bp1tm}True masses.}
%\end{table}

\subsubsection{Endpoint values}
\begin{table}[h]
\begin{center}
\begin{tabular}{|c|c|c|c|c|}
\hline
 & $m_{ll}^{max}$ & $m_{jll}^{max}$ & $m_{jl(lo)}^{max}$ & $m_{jl(hi)}^{max}$ \\
\hline
Expected & 109.9 & 528.5 & 336.8 & 451.8 \\
\hline
Fit & $109.7\pm 0.4$ & $533.3\pm 3.0$ & $341.2\pm 6.0$ & $457.1\pm 2.4$ \\
\hline
\end{tabular}
\end{center}
%\caption{Endpoint values.}
\label{table:bp1ep}
\end{table}

\subsubsection{Mass solutions}
\begin{table}[h]
\begin{center}
\begin{tabular}{|c|c|c|c|c|}
\hline
Region~\cite{GMO} & $m_{\tilde l}-m_{\tilde\chi_1^0}$ & $m_{\tilde\chi_2^0}-m_{\tilde l}$ & $m_{\tilde q}-m_{\tilde\chi_2^0}$ & $m_{\tilde q}$ \\
\hline
Expected & 37.9 & 73.5 & 429 & 660 \\
\hline
(1,1) & $73.0\pm 1.7$ & $43.1\pm 1.0$ & $459\pm 11$ & $876\pm 62$ \\
(1,2) & \multicolumn{4}{|c|}{Imaginary} \\
{\bf (1,3)} & ${\bf 37.7\pm 1.1}$ & ${\bf 73.3\pm 1.9}$ & ${\bf 433\pm 7}$ & ${\bf 655\pm 35}$ \\
\hline
(4,1) & \multicolumn{4}{|c|}{Imaginary} \\
(4,2) & $33.3\pm 3.5$ & $76.5\pm 3.0$ & $423\pm 3.2$ & $551\pm 5$\\
(4,3) & $37.2\pm 1.3$ & $73.5\pm 2.0$ & $423\pm 3.0$ & $621\pm 15$\\
\hline
\end{tabular}
\end{center}
%\caption{}
\label{table:bp1ms}
%\caption{Mass solutions. Correct solution is in bold.}
\end{table}
Correct solution is in bold.

\begin{figure}[h]
\begin{center}
\includegraphics[width=\columnwidth]{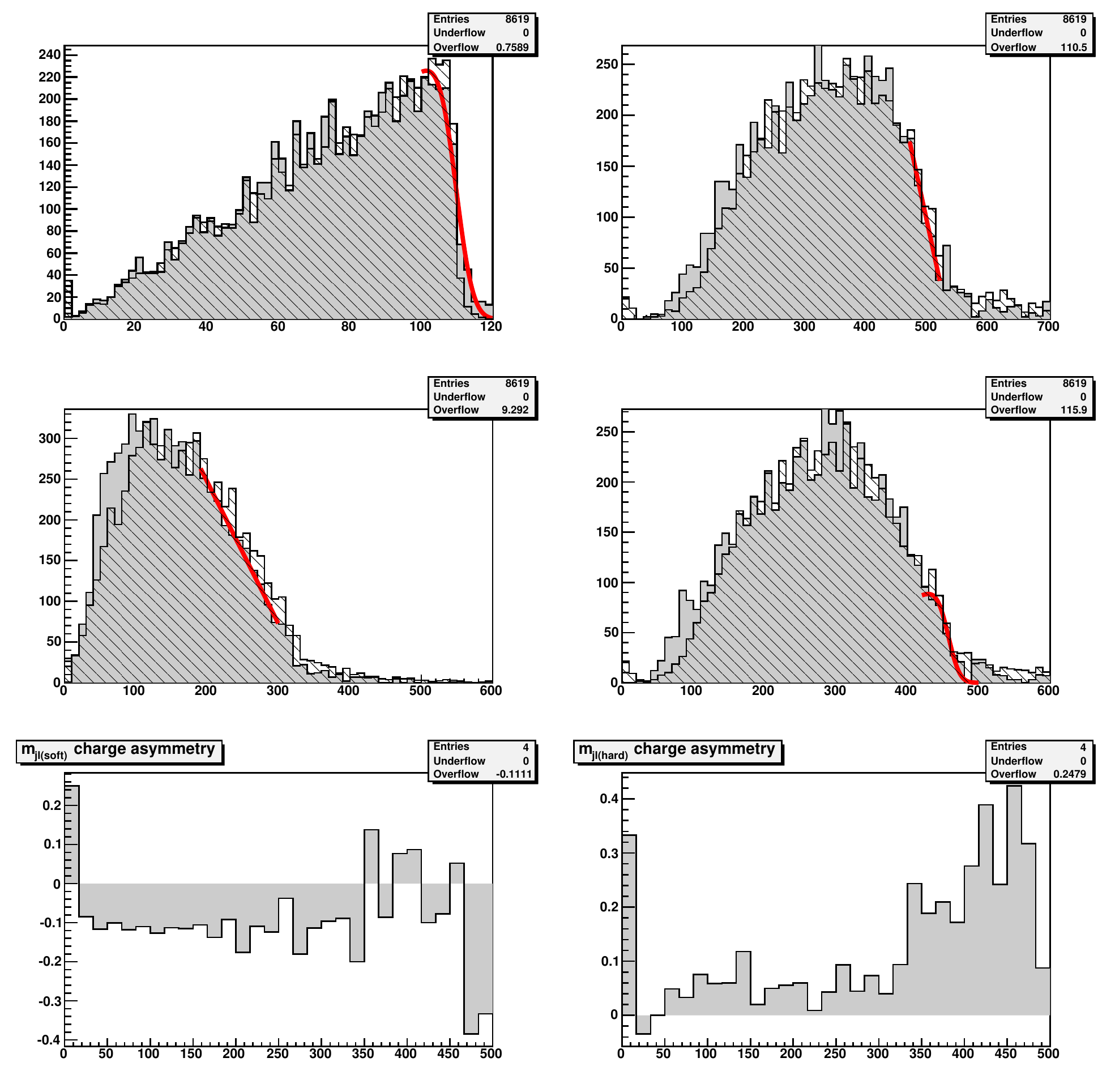}
\end{center}
\caption{BP~1 invariant mass distributions and endpoint fits. Upper left, $m_{ll}$; upper right, $m_{jll}$; middle left, $m_{jl(lo)}$; middle right, $m_{jl(hi)}$.  Cross-hatched region is parton-level, and solid region is jet-level.  Lower left, $m_{jl^+(soft)}-m_{jl^-(soft)}$; lower right, $m_{jl^+(hard)}-m_{jl^-(hard)}$.}
\label{fig:bp1}
\end{figure}

\subsubsection{Notable branching ratios}
\begin{table}[h]
\begin{center}
\begin{tabular}{|c|c|c|}
\hline
Parent & Daughters & Branching ratio \\
\hline
$\tilde u_L$ & $u_L\tilde\chi_2^0$ & 0.33 \\
& $d_L\tilde\chi_1^\pm$ & 0.66 \\
\hline
$\tilde u_R$ & $u_R\tilde\chi_1^0$ & 1.0 \\
& $u_R\tilde\chi_2^0$ & $1.1\times 10^{-3}$ \\
\hline
$\tilde\chi_2^0$ & $l_L^\pm\tilde l_L^\mp$ & 0.42 \\
& $l_R^\pm\tilde l_R^\mp$ & $5.8\times 10^{-5}$\\
& $\nu_L\tilde\nu_L$ & 0.58 \\
\hline
\end{tabular}
\end{center}
\label{table:bp1br}
%\caption{Notable branching ratios.}
\end{table}

\clearpage
\subsection{Benchmark point 2 (BP 2)}

\subsubsection{True masses}
$$m_{\tilde\chi_1^0}=120.1,\quad m_{\tilde l}=193.7,\quad m_{\tilde\chi_2^0}=234.5,\quad m_{\tilde q}=643.5$$
\subsubsection{Endpoint values}
\begin{table}[h]
\begin{center}
\begin{tabular}{|c|c|c|c|c|}
\hline
 & $m_{ll}^{max}$ & $m_{jll}^{max}$ & $m_{jl(lo)}^{max}$ & $m_{jl(hi)}^{max}$ \\
\hline
Expected & 103.7 & 514.6 & 337.8 & 470.0 \\
\hline
Fit & $103.7\pm 0.3$ & $516.7\pm 1.2$ & $338.0\pm 1.2$ & $468.5\pm 1.9$ \\
\hline
\end{tabular}
\end{center}
\label{table:bp2ep}
\end{table}
A fit to the $m_{jl(hi)}$ endpoint with a vertical edge gives $m_{jl(hi)}^{max}=469.1\pm 4.8$.  

For calculating charge asymmetry, $m_{jl(near)}=349.5\pm 2.8$

\subsubsection{Mass solutions}
\begin{table}[h]
\begin{center}
\begin{tabular}{|c|c|c|c|c|}
\hline
Region~\cite{GMO} & $m_{\tilde l}-m_{\tilde\chi_1^0}$ & $m_{\tilde\chi_2^0}-m_{\tilde l}$ & $m_{\tilde q}-m_{\tilde\chi_2^0}$ & $m_{\tilde q}$ \\
\hline
Expected & 73.5 & 40.8 & 409 & 643 \\
\hline
{\bf (1,1)} & ${\bf 72.9\pm 0.7}$ & ${\bf 40.8\pm 0.3}$ & ${\bf 414\pm 3}$ & ${\bf 660\pm 10}$ \\
(1,2) & \multicolumn{4}{|c|}{Imaginary} \\
(1,3) & $30.0\pm 0.7$ & $75.0\pm 0.8$ & $412\pm 1.4$ & $563\pm 4.6$ \\
\hline
(4,1) & \multicolumn{4}{|c|}{Imaginary} \\
(4,2) & $23.7\pm 1.2$ & $80.0\pm 1.2$ & $413\pm 1$ & $528\pm 2$\\
(4,3) & $30.0\pm 0.7$ & $74.8\pm 0.8$ & $412\pm 1$ & $562\pm 4$\\
\hline
\end{tabular}
\end{center}
\label{table:bp2ms}
\end{table}
Correct solution is in bold.

\begin{figure}
\begin{center}
\includegraphics[width=\columnwidth]{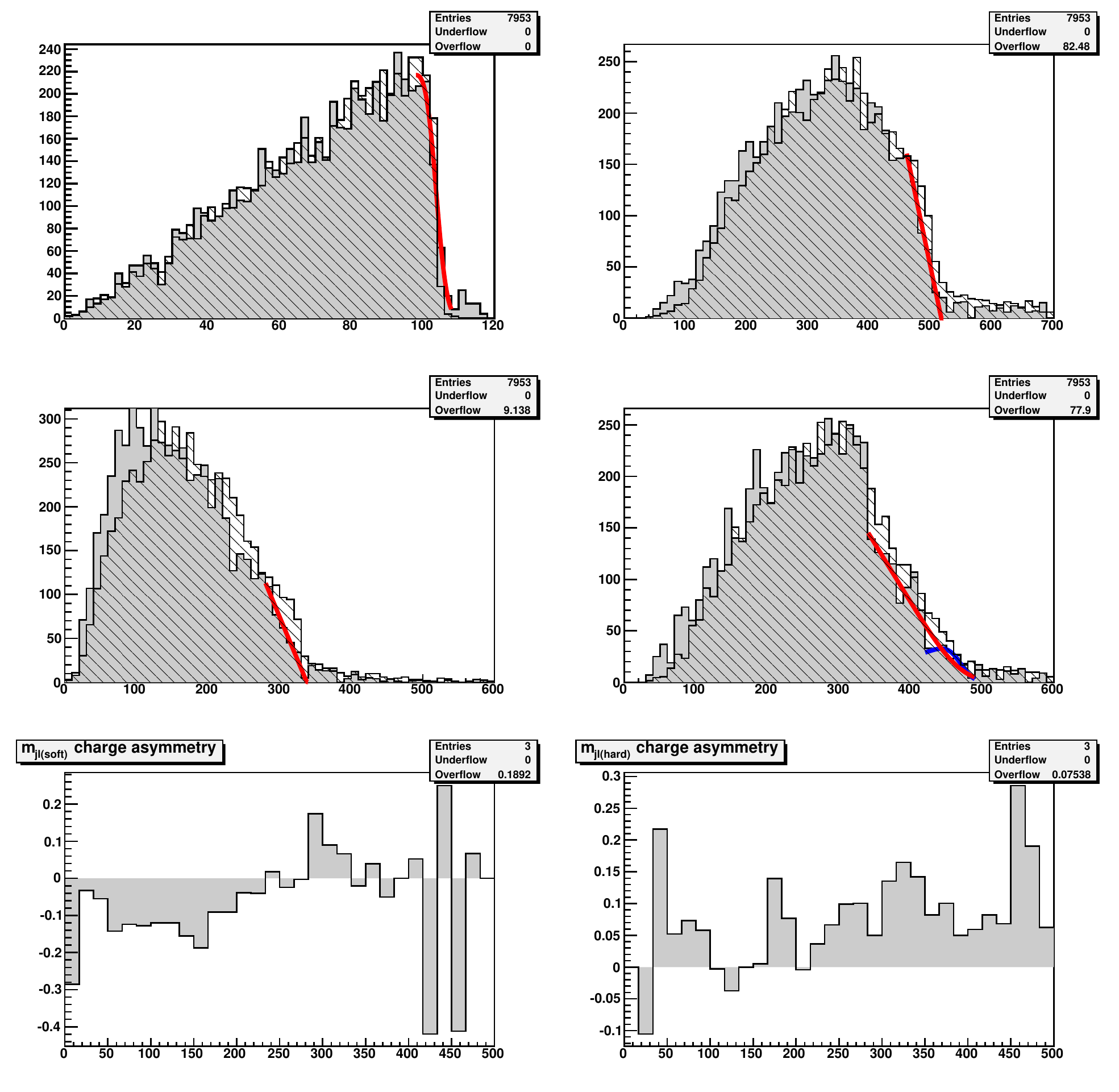}
\end{center}
\caption{BP~2 invariant mass distributions and endpoint fits. Upper left, $m_{ll}$; upper right, $m_{jll}$; middle left, $m_{jl(lo)}$; middle right, $m_{jl(hi)}$.  For $m_{jl(hi)}$, the red curve is the fit with a ramp function, and the blue curve is the fit with a vertical edge.  Cross-hatched region is parton-level, and solid region is jet-level.  Lower left, $m_{jl^+(soft)}-m_{jl^-(soft)}$; lower right, $m_{jl^+(hard)}-m_{jl^-(hard)}$.}
\label{fig:bp2}
\end{figure}

\clearpage
\subsubsection{Notable branching ratios}
\begin{table}[h]
\begin{center}
\begin{tabular}{|c|c|c|}
\hline
Parent & Daughters & Branching ratio \\
\hline
$\tilde u_L$ & $u_L\tilde\chi_2^0$ & 0.33 \\
& $d_L\tilde\chi_1^\pm$ & 0.66 \\
\hline
$\tilde u_R$ & $u_R\tilde\chi_1^0$ & 1.0 \\
& $u_R\tilde\chi_2^0$ & $5.0\times 10^{-4}$ \\
\hline
$\tilde\chi_2^0$ & $l_L^\pm\tilde l_L^\mp$ & 0.36 \\
& $l_R^\pm\tilde l_R^\mp$ & $5.7\times 10^{-4}$\\
& $\nu_L\tilde\nu_L$ & 0.64 \\
\hline
\end{tabular}
\end{center}
\label{table:bp2br}
\end{table}

\clearpage
\subsection{Benchmark point 3 (BP 3)}

\subsubsection{True masses}
$$m_{\tilde\chi_1^0}=106.4,\quad m_{\tilde l}=135.3,\quad m_{\tilde\chi_2^0}=166.9,\quad m_{\tilde q}=639.2$$
\subsubsection{Endpoint values}
\begin{table}[h]
\begin{center}
\begin{tabular}{|c|c|c|c|c|}
\hline
 & $m_{ll}^{max}$ & $m_{jll}^{max}$ & $m_{jl(lo)}^{max}$ & $m_{jl(hi)}^{max}$ \\
\hline
Expected & 60.4 & 475.6 & 324.4 & 381.4 \\
\hline
Fit & $60.4\pm 0.1$ & $473.2\pm 4.5$ & $323.3\pm 0.9$ & $370.2\pm 4.4$ \\
\hline
\end{tabular}
\end{center}
\label{table:bp3ep}
\end{table}

\begin{table}[h]
\begin{center}
\begin{tabular}{|c|c|}
\hline
 Secondary edge & $m_{ll}^{max}$\\
 \hline
 $\tilde\chi_2^\pm\rightarrow\tilde\nu_L l^\pm\rightarrow\tilde\chi_1^\pm ll$ &  93.0\\
$\tilde\chi_4^0\rightarrow\tilde l_L l\rightarrow\tilde\chi_2^0 ll $ & 107.4\\
$\tilde\chi_4^0\rightarrow\tilde l_L l\rightarrow\tilde\chi_1^0 ll $ & 176.4\\\hline

\hline
\end{tabular}
\end{center}
\label{table:bp3ep2}
\end{table}

\subsubsection{Mass solutions}
\begin{table}[h]
\begin{center}
\begin{tabular}{|c|c|c|c|c|}
\hline
Region~\cite{GMO} & $m_{\tilde l}-m_{\tilde\chi_1^0}$ & $m_{\tilde\chi_2^0}-m_{\tilde l}$ & $m_{\tilde q}-m_{\tilde\chi_2^0}$ & $m_{\tilde q}$ \\
\hline
Expected & 28.9 & 31.6 & 472 & 639\\
\hline
(1,1) & $33.8\pm 0.4$ & $27.3\pm 0.3$ & $580\pm 48$ & $988\pm 170$ \\
{\bf (1,2)} & ${\bf 29.1\pm 1.0}$ & ${\bf 31.4\pm 0.9}$ & ${\bf 491\pm 12}$ & ${\bf 693\pm 31}$\\
(1,3) & $28.7\pm 1.0$ & $32.1\pm 0.6$ & $437\pm 24$ & $558\pm 50$ \\
\hline
(4,1) & \multicolumn{4}{|c|}{Imaginary} \\
(4,2) & $38.1\pm 2.2$ & $30.1\pm 0.6$ & $404\pm 6$ & $480\pm 7$\\
(4,3) & $30.6\pm 1.0$ & $31.5\pm 0.6$ & $411\pm 5$ & $506\pm 10$\\
\hline
\end{tabular}
\end{center}
\label{table:bp3ms}
\end{table}
Correct solution is in bold.

\begin{figure}
\begin{center}
\includegraphics[width=\columnwidth]{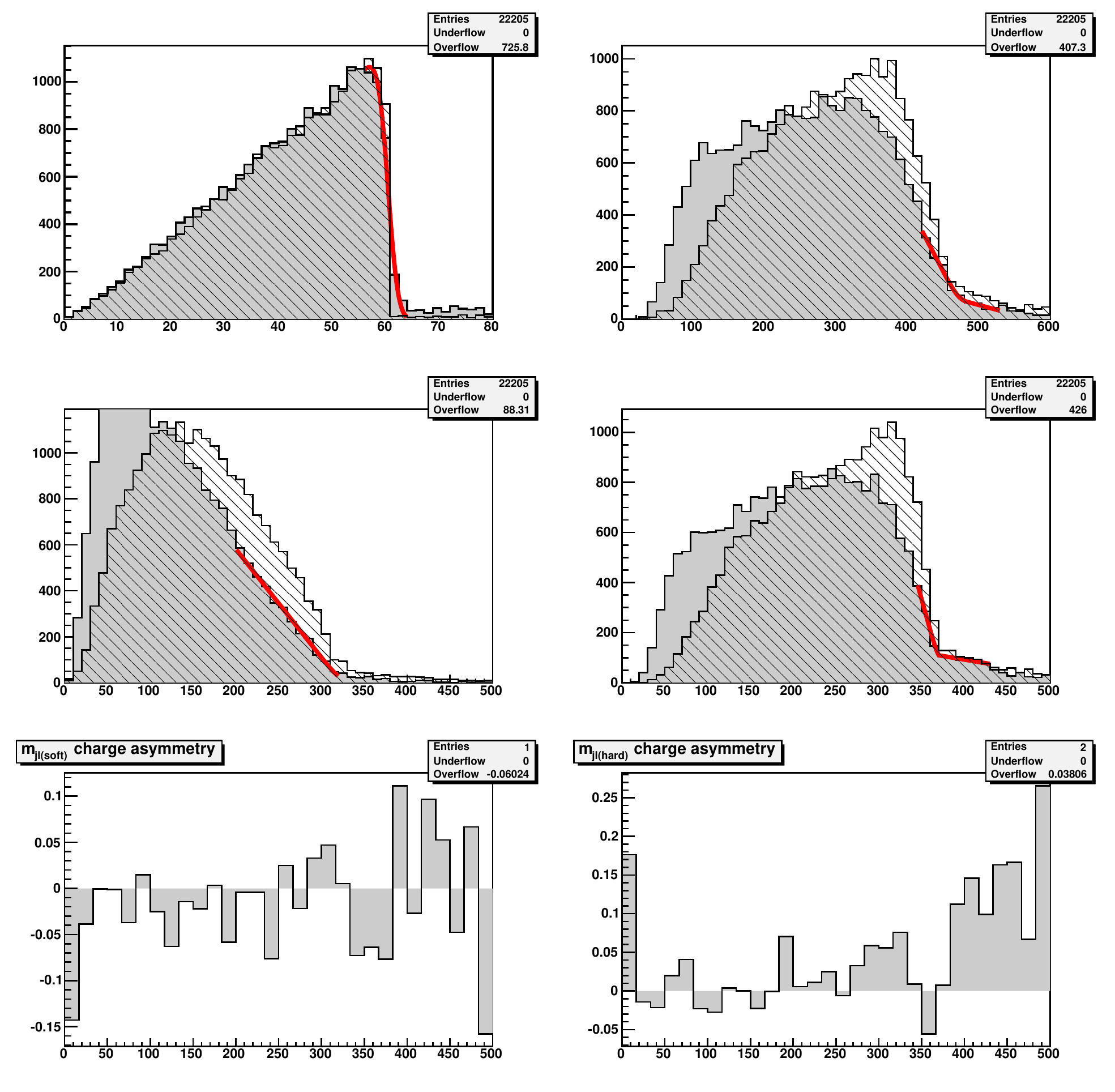}
\end{center}
\caption{BP~3 invariant mass distributions and endpoint fits. Upper left, $m_{ll}$; upper right, $m_{jll}$; middle left, $m_{jl(lo)}$; middle right, $m_{jl(hi)}$.  Cross-hatched region is parton-level, and solid region is jet-level.  Lower left, $m_{jl^+(soft)}-m_{jl^-(soft)}$; lower right, $m_{jl^+(hard)}-m_{jl^-(hard)}$.}
\label{fig:bp3}
\end{figure}

\clearpage
\subsubsection{Notable branching ratios}
\begin{table}[h]
\begin{center}
\begin{tabular}{|c|c|c|}
\hline
Parent & Daughters & Branching ratio \\
\hline
$\tilde g$ & $q\tilde q_L$ & 0.40 \\
& $q\tilde q_R$ & 0.59\\
\hline
$\tilde u_L$ & $u_L\tilde\chi_2^0$ & 0.17 \\
& $u_L\tilde\chi_4^0$ & 0.16 \\
& $d_L\tilde\chi_1^\pm$ & 0.37 \\
& $d_L\tilde\chi_2^\pm$ & 0.30 \\
\hline
$\tilde d_L$ & $d_L\tilde\chi_2^0$ & 0.095 \\
& $d_L\tilde\chi_4^0$ & 0.19 \\
& $u_L\tilde\chi_1^\pm$ & 0.20 \\
& $u_L\tilde\chi_2^\pm$ & 0.45 \\
\hline
$\tilde u_R$ & $u_R\tilde\chi_1^0$ & 0.76 \\
& $u_R\tilde\chi_2^0$ & 0.22 \\
\hline
$\tilde\chi_2^0$ & $l_R^\pm\tilde l_R^\mp$ & 1.0\\
\hline
$\tilde\chi_4^0$ & $l_L^\pm\tilde l_L^\mp$ & 0.21\\
& $l_R^\pm\tilde l_R^\mp$ & 0.02\\
& $\nu_L\tilde\nu_L$ & 0.4\\
\hline
$\tilde\chi_1^\pm$ & $l_L^\pm \nu_L \tilde\chi_1^0$ & 0.36 \\
\hline
$\tilde\chi_2^\pm$ & $l_L^\pm \tilde\nu_L $ & 0.33 \\
\hline
\end{tabular}
\end{center}
\label{table:bp3br}
\end{table}

\clearpage
\subsection{Comparison point (CP)}

\subsubsection{True masses}
$$m_{\tilde\chi_1^0}=118.8,\quad m_{\tilde l}=144.5,\quad m_{\tilde\chi_2^0}=222.0,\quad m_{\tilde q}=666$$
\subsubsection{Endpoint values}
\begin{table}[h]
\begin{center}
\begin{tabular}{|c|c|c|c|c|}
\hline
 & $m_{ll}^{max}$ & $m_{jll}^{max}$ & $m_{jl(lo)}^{max}$ & $m_{jl(hi)}^{max}$ \\
\hline
Expected & 95.9 & 530.3 & 310.4 & 476.6 \\
\hline
Fit & $95.3\pm 0.4$ & $547.6\pm 6.2$ & $309.0\pm 2.7$ & $479.6\pm 15.5$ \\
\hline
\end{tabular}
\end{center}
\label{table:cpep}
\end{table}

\subsubsection{Mass solutions}
\begin{table}[h]
\begin{center}
\begin{tabular}{|c|c|c|c|c|}
\hline
Region~\cite{GMO} & $m_{\tilde l}-m_{\tilde\chi_1^0}$ & $m_{\tilde\chi_2^0}-m_{\tilde l}$ & $m_{\tilde q}-m_{\tilde\chi_2^0}$ & $m_{\tilde q}$ \\
\hline
Expected & 25.7 & 77.5 & 444 & 666 \\
\hline
(1,1) & $75.2\pm 3.8$ & $31.6\pm 0.7$ & $526\pm 67$ & $1080\pm 400$ \\
(1,2) & \multicolumn{4}{|c|}{Imaginary} \\
{\bf (1,3)} & ${\bf 27.6\pm 3.3}$ & ${\bf 76.0\pm 4.5}$ & ${\bf 497\pm 51}$ & ${\bf 848\pm 236}$ \\
\hline
(4,1) & \multicolumn{4}{|c|}{Imaginary} \\
(4,2) & $14.9\pm 3.8$ & $89.0\pm 7.9$ & $444\pm 7$ & $574\pm 10$\\
(4,3) & $24.9\pm 2.5$ & $78.3\pm 45.3$ & $444\pm 6$ & $664\pm 30$\\
\hline
\end{tabular}
\end{center}
\label{table:cpms}
\end{table}
Correct solution is in bold.

\begin{figure}
\begin{center}
\includegraphics[width=\columnwidth]{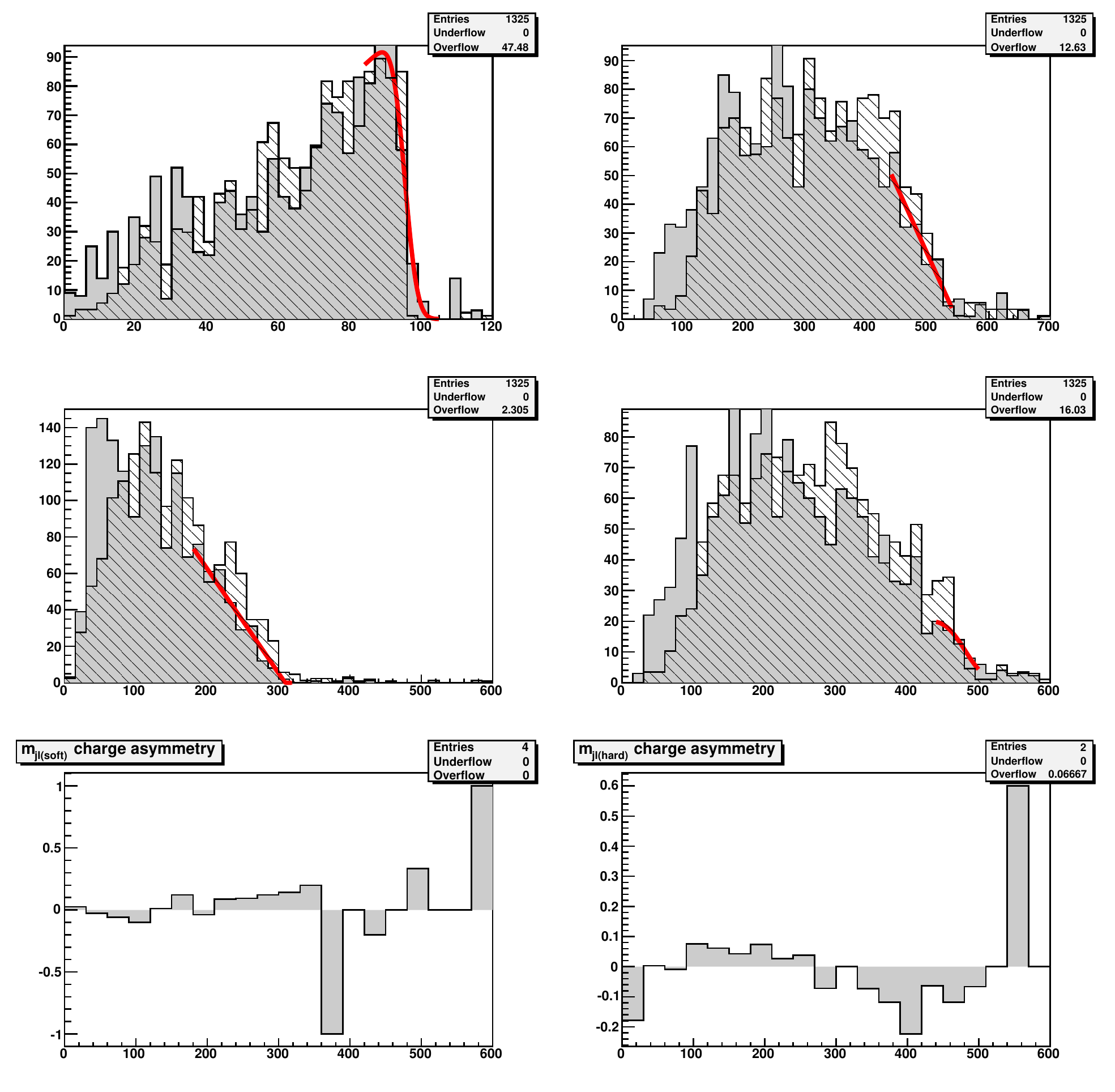}
\end{center}
\caption{Comparison point invariant mass distributions and endpoint fits. Upper left, $m_{ll}$; upper right, $m_{jll}$; middle left, $m_{jl(lo)}$; middle right, $m_{jl(hi)}$.  Cross-hatched region is parton-level, and solid region is jet-level.  Lower left, $m_{jl^+(soft)}-m_{jl^-(soft)}$; lower right, $m_{jl^+(hard)}-m_{jl^-(hard)}$.}
\label{fig:cp}
\end{figure}

\subsubsection{Notable branching ratios}
\begin{table}[h]
\begin{center}
\begin{tabular}{|c|c|c|}
\hline
Parent & Daughters & Branching ratio \\
\hline
$\tilde u_L$ & $u_L\tilde\chi_2^0$ & 0.32 \\
& $d_L\tilde\chi_1^\pm$ & 0.65 \\
\hline
$\tilde u_R$ & $u_R\tilde\chi_1^0$ & 0.99 \\
& $u_R\tilde\chi_2^0$ & $4.7\times 10^{-3}$ \\
\hline
$\tilde\chi_2^0$ & $l_R^\pm\tilde l_R^\mp$ & $0.052$\\
& $\tau^\pm\tilde\tau_1^\mp$ & $0.44$\\
& $\nu_L\tilde\nu_L$ & 0.50 \\
\hline
\end{tabular}
\end{center}
\label{table:cpbr}
\end{table}

\end{document}